\documentclass{ptephy_v1}

\usepackage{amsmath,amssymb}
\usepackage{subfig}
\usepackage{url}
\usepackage{tikz}
\usepackage{float}
\usepackage{braket}

\newcommand{\OD}[1]{\frac{\mathrm{d}}{\mathrm{d} #1}}
\newcommand{\OOD}[1]{\frac{\mathrm{d}^2}{\mathrm{d} #1^2}}
\newcommand{\ODD}[2]{\frac{\mathrm{d} #1}{\mathrm{d} #2}}

\newcommand{\PD}[1]{\frac{\partial}{\partial #1}}

\newcommand{\PDD}[2]{\frac{\partial #1}{\partial #2}}

\renewcommand{\Re}[1]{\mathrm{Re}\left[ #1 \right]}
\renewcommand{\Im}[1]{\mathrm{Im}\left[ #1 \right]}

\newcommand{\cons}{{\rm cons}}
\newcommand{\diss}{{\rm diss}}
\newcommand{\baligned}{}
\newcommand{\ealigned}{}
\begin{document}

\title{Renormalization group analysis of superradiant growth of self-interacting axion cloud}


\author[1,*]{Hidetoshi Omiya}
\affil{Department of Physics, Kyoto University, Kyoto 606-8502, Japan \email{omiya@tap.scphys.kyoto-u.ac.jp}}

\author[1]{Takuya Takahashi}

\author[1,2]{Takahiro Tanaka}
\affil{Center for Gravitational Physics, Yukawa Institute for Theoretical Physics, Kyoto University, Kyoto 606-8502, Japan}


\begin{abstract}%
There are strong interests in considering ultra-light scalar fields (especially axion) around a rapidly rotating black hole because of the possibility of observing gravitational waves from axion condensate (axion cloud) around black holes. Motivated by this consideration, we propose a new method to study the dynamics of an ultra-light scalar field with self-interaction around a rapidly rotating black hole, which uses the dynamical renormalization group method. We find that for relativistic clouds, saturation of the superradiant instability by the scattering of the axion due to the self-interaction does not occur in the weakly non-linear regime when we consider the adiabatic growth of the cloud from a single superradiant mode. This may suggest that for relativistic axion clouds, an explosive phenomenon called the Bosenova may inevitably happen, at least once in its evolutionary history.
\end{abstract}

\maketitle

\section{Introduction}
String Theory provides us with an interesting scenario that our universe contains plenty of ultralight axion-like particles, which is called {\it axiverse}~\cite{Arvanitaki:2009fg}. Detecting particles of this type will provide us with a lot of information about physics beyond the Standard Model, or even a clue to the string theory. In Ref.~\cite{Arvanitaki:2009fg}, the authors proposed that these axion-like particles may cause observable astrophysical or cosmological phenomena, such as the polarization of CMB or the step-like features in the matter power spectrum. In this paper, we will concentrate on the other possible phenomena involving a black hole.

We consider a massive scalar field around a Kerr black hole. As is well known, a rotating black hole possesses the ergoregion. If the negative energy field excitations allowed in the ergoregion fall into the black hole, energy conservation implies that the external field gains energy from the black hole. This is the {\it superradiance}, {\it i.e.}, energy extraction from a black hole via waves, analogous to the famous Penrose process~\cite{Brito:2015oca}.

Now, recall that the scalar field is massive. The mass makes the scalar field bounded around the black hole. This bound state continues to gain energy by the superradiance and keeps growing, which is called {\it superradiant instability}. Because we mainly concern QCD axion or axion-like particles from string theory in this paper, we refer to such an unstable bound state as an {\it axion cloud}~\cite{Arvanitaki:2009fg, Arvanitaki:2010sy}. 

The growth rate of an axion cloud was calculated approximately in Ref.~\cite{Zouros:1979iw, Detweiler:1980uk} and numerically in Ref.~\cite{Dolan:2007mj}. These works show that the largest growth rate is $\omega_I \sim10^{-7}M^{-1}$ in units of $G = c = \hbar =1$, where $M$ is the mass of the black hole. $\omega_I$ can be understood as the imaginary part of the scalar field oscillation frequency $\omega$. The largest growth rate is achieved when the effective gravitational coupling governed by $\alpha_g = \mu M$ is $\sim 1$, with $\mu$ being the mass of the axion field. For example, if we consider a solar mass black hole, $\mu \sim 10^{-10}$eV gives the largest growth rate. For the superradiant instability to be efficient within the age of the Universe, the mass of the scalar field has to be in the range $10^{-20} \sim 10^{-10}$eV.  Axion-like particles motivated by string theory naturally have a mass in this range~\cite{Svrcek:2006yi}.

If ultralight particles beyond the Standard Model such as an axion exist, an axion cloud forms around a black hole and interesting phenomena happen. Similar to the photon emission from a hydrogen atom, the axion cloud can emit gravitational waves by the transition between different energy levels. Also, axion pair annihilation can result in gravitational wave emission~\cite{Arvanitaki:2010sy, Yoshino:2013ofa}. Another phenomenon is the loss of spin and energy of the black hole as the axion cloud grows. As a result, a characteristic feature in black hole spin and mass distribution is expected~\cite{Arvanitaki:2010sy, Brito:2014wla}. 

Including the self-interaction of axion can have dramatic effects on the axion cloud. The self-interaction of the axion is typically attractive, and thus when an axion cloud gets heavy enough, this attractive force would lead to a collapse of the axion cloud. This collapse is called {\it bosenova} and a burst of gravitational waves is expected during the collapse~\cite{Arvanitaki:2010sy, Yoshino:2012kn, Yoshino:2015nsa}. These phenomena have attracted great interest owing to the possibility of their detection~\cite{Arvanitaki:2014wva, Brito:2017zvb}. To prepare for the future detection of axion clouds through observations of gravitational waves or astronomical observations of black hole spins, precise analysis of the evolution of axion clouds is important.

To do so, we must correctly take into account the self-interaction of the axion. Only a few works have discussed the self-interaction incisively. Numerical simulation~\cite{Yoshino:2012kn,Yoshino:2015nsa} and analytic treatment within the non-relativistic approximation~\cite{Arvanitaki:2010sy} were made to show that the self-interaction causes the bosenova. However, in the numerical simulations, there is a difficulty due to the huge hierarchy of time scales between the growth rate $\omega_I$ and the dynamical time scale $\omega_R$. In Refs.~\cite{Yoshino:2012kn,Yoshino:2015nsa}, they used a large amplitude axion cloud as the initial condition to overcome this problem. However, in a realistic scenario, an axion cloud starts to evolve with a small amplitude seeded by fluctuation. To answer the question of whether bosenova occurs in a realistic case, adopting a particular configuration with a large amplitude as the initial condition would be difficult to justify. Also, the most interesting case, in which the growth rate is maximized, is in the relativistic regime, so that the non-relativistic approximation would not be satisfactory. 

Apart from the bosenova, the self-interaction has one more important effect: the energy loss of the axion cloud due to the scattering of the axion. Using the order estimate in Ref.~\cite{Arvanitaki:2010sy}, the energy loss due to the self-interaction is typically more important than that due to the gravitational wave emission when the axion cloud grows by superradiant instability, but this effect is usually ignored in the literature. Our naive expectation is that when the cloud gets denser, this energy loss may balance the superradiant instability. This is because the time scale of the energy gain scales as $\propto |A|^{0}$, while the energy loss scales as $\propto |A|^{-4}$, where $|A|$ is the amplitude of the cloud. Thus, superradiant growth of the cloud should terminate in the weakly non-linear regime, and no explosive phenomena induced by the strong non-linear effects should happen.

Our main focus of this paper is to clarify the effects of self-interaction on the dynamics of an axion cloud avoiding the drawbacks in numerical simulation and non-relativistic approximation. We use perturbation theory to tackle the problem of axion self-interaction. When we apply the perturbation theory to a non-linear problem, we often encounter a secular term that destroys the validity of the perturbation theory. We encounter the same difficulty in the problem that we are concerned with. To overcome this difficulty, we use the renormalization group (RG) method for differential equations ~\cite{Chen:1994zza}. 

Using the RG method, we obtain the evolution equation that describes the long term behavior of the axion cloud composed of a single superradiant mode. Our main conclusion is that the axion self-interaction would not terminate the superradiant instability by the energy loss within the validity range of our approximation, contrary to our naive initial expectation. What we will find is that the non-linearity accelerates the superradiant instability of the axion cloud because of the attractive nature of axion self-interaction.

This paper is organized as follows. In Sec.~\ref{section:2}, we review superradiant instability and how a growing cloud around a rotating black hole is formed. In Sec.~\ref{section:3} we derive the evolution equation of the axion cloud by the RG method. In Sec.~\ref{section:4} we analyze the evolution of the axion cloud by the equations derived in Sec.~\ref{section:3}. In Sec.~\ref{section:add}, we make some comments on the recent paper \cite{Baryakhtar:2020gao}, whose claim may look conflicting with ours. Finally, we summarize our result and discuss the implication of our result in Sec.~\ref{section:5}. Since the RG method is important for our analysis, we provide two simple examples of the RG method for differential equations in App.\ref{App:A} that may help understanding the application of the RG method to the current problem.
In this paper, we use $G=c=\hbar = 1$ units, unless otherwise stated.

\section{Superradiant instability}
\label{section:2}

In this section, we review the superradiant instability of an axion field around a Kerr black hole. For a recent review on this topic, see Ref.~\cite{Brito:2015oca}.

We consider the following action for an axion field $\phi$:
\begin{align}\label{action}
	S = \int d^4 x \sqrt{-g} \left\{ - \frac{1}{2}g^{\mu\nu}\partial_\mu \phi \partial_\nu \phi - \mu^2 F_a^2 \left(1 - \cos\left(\frac{\phi}{F_a}\right)\right)\right\}~.
\end{align}
Here, $F_a$ is the decay constant of axion and $g_{\mu\nu}$ is the metric of the Kerr spacetime, which gives the line element
\begin{align}
\label{kerrmetric}
	ds^2 &=  - \left(1 - \frac{2 M r}{\rho^2}\right)dt^2 - \frac{4 a M r\sin^2\theta}{\rho^2} dt d\varphi \cr
	 &\ \ \ \ \ + \left[(r^2 + a^2)\sin^2\theta+ \frac{2 M r}{\rho^2} a^2 \sin^4 \theta \right]d\varphi^2 + \frac{\rho^2}{\Delta} dr^2 + \rho^2 d\theta^2~,
\end{align}
with 
\begin{align}
	\Delta &= r^2 - 2 Mr + a^2~, & \rho^2 &= r^2 + a^2\cos^2\theta~.
\end{align}
There are two horizons in the Kerr spacetime specified by the solutions of $\Delta = 0$, which are
\begin{align}
	r_\pm = M \pm \sqrt{M^2 - a^2}~.
\end{align}
As we are interested in axions, we adopt the cosine type potential induced by the quantum effect~\cite{Weinberg:1996kr}. From the action \eqref{action}, the equation of motion for the axion field is
\begin{equation}\label{eom1}
	\left(\square_g \phi - \mu^2 F_a \sin\left(\frac{\phi}{F_a}\right)\right)= 0~.
\end{equation}
Here, $\square_g$ is the d'Alembertian on the Kerr metric. 

In this section, we solve the linearized version of Eq.~\eqref{eom1}, which is
\begin{align}
	\left(\square_g  - \mu^2 \right)\phi = 0~.
\end{align}
In the Kerr spacetime, separation of variables is possible~\cite{Brill:1972xj}. We take
\begin{align}
	\label{sep}
	\phi = \Re{e^{-i(\omega t - m \varphi)}S_{lm\omega}(\theta)R_{lm\omega}(r)}\,,
\end{align}
as an ansatz for the axion field, to get
\begin{gather}
		\label{angularEOM}
		\frac{1}{\sin\theta}\OD{\theta}\left(\sin\theta \ODD{{S}_{lm\omega}}{\theta}\right) + [c^2(\omega) \cos^2\theta - \frac{m^2}{\sin^2\theta}]S_{lm\omega} = - \Lambda_{lm}(\omega) S_{lm\omega}~,\\
		\label{radialEOM}
		\OD{r}\left(\Delta\ODD{R_{lm\omega}}{r}\right)  + \left[\frac{K^2(\omega)}{\Delta} - \mu^2 r^2 -\lambda_{lm}(\omega) \right]R_{lm\omega} = 0~,
\end{gather}
where
\begin{align}
\baligned{}
		 c^2 (\omega) &= a^2 (\omega^2 - \mu^2)~, \qquad K(\omega) = (r^2+a^2)\omega - am~,
\ealigned{}\cr
	 \lambda_{lm}(\omega) & = -2am M \omega +a^2\omega^2 +\Lambda_{lm}(\omega)~.
\end{align}
$\Lambda_{lm}(\omega)$ is the separation constant and has to be calculated numerically. Also, we normalize the angular solution as 
\begin{align}
	\int d\cos\theta \ S_{lm\omega}(\theta)^2 = 1~.
\end{align}

Following Refs.~\cite{Zouros:1979iw} and~\cite{Dolan:2007mj}, we can show that this system develops instability, {\it i.e.}, $\omega_I = \Im{\omega} > 0$. In the Kerr spacetime, there is a timelike Killing vector $\xi^\mu \equiv (\partial_t)^{\mu}$. Using the conservation of $T^\mu_{~\nu}$ and the Killing equation, we obtain
\begin{align}\label{eq:13}
		(T^\mu_{~\nu} \xi^\nu)_{;\mu} = 0~.
\end{align}
Here, we introduce the ingoing Kerr coordinates $(\tilde{t},r,\theta,\tilde{\varphi})$ defined by 
\begin{align}
	d\tilde{t} &= dt + \frac{r^2 + a^2}{\Delta}dr~, & d\tilde{\varphi} = d\varphi + \frac{a}{\Delta} dr~.
\end{align}
Integrating Eq.~\eqref{eq:13} over $\tilde{t}=$constant surface $\Sigma$, we obtain
\begin{align}
	\label{Gauss}
	-\PD{\tilde{t}} \left(\int_\Sigma \sqrt{-\tilde{g}^{00}}\rho^2 \sin\theta dr d\theta d\tilde{\varphi}\  {T^{\mu}}_\nu \xi^\nu n_\mu \right)= \int_{\partial \Sigma}\rho^2 \sin\theta d\theta d\tilde{\varphi}\  {T^{\mu}}_\nu \xi^\nu k_\mu ~,
\end{align}
where $n_{\mu}=-\delta^0_\mu$ and $k_\mu = - \delta^1_\mu$. By substituting the ansatz \eqref{sep} into Eq.~\eqref{Gauss}, we find
\begin{align}
	\label{energyintegral}
	2 \omega_I \int_\Sigma \sqrt{-\tilde{g}^{00}}\rho^2 \sin\theta dr d\theta d\tilde{\varphi}\ {T^0}_0 = -2 M r_+ 2\pi|R_{lm\omega}(r_+)|^2 (|\omega|^2 - m \Omega_H \omega_R)~,
\end{align}
where $\Omega_H$ is the angular velocity of the outer horizon given by $\Omega_H \equiv a/2Mr_+$. Note that contribution to the right hand side from the boundary at $r= \infty$ disappears because we are considering a bound state, which vanishes exponentially there. We refer to the real part of $\omega$ as $\omega_R$. Equation \eqref{energyintegral} shows that the instability occurs when 
\begin{align}
	\frac{|\omega|^2}{\omega_R} < m \Omega_H
\end{align}
holds. Because the condition for the instability to occur is the same as the superradiance condition $\omega < m \Omega_H$ if $\omega_I \ll \omega_R$, we will denote this unstable mode as the superradiant mode. The superradiant mode grows exponentially around the black hole to develop a condensate of the axion field, which is called an axion cloud. 

There are several papers on the evaluation of $\omega_I$. In Ref.~\cite{Zouros:1979iw}, the WKB method was used with the assumption $\mu M \gg 1$ and in Ref.~\cite{Detweiler:1980uk}, the matched asymptotic expansion method was used with $ \mu M \ll 1$. These works show that $\omega_I$ is the largest in the parameter region with $\mu M \sim 1$. In this regime, $\omega_I$ has to be calculated numerically~\cite{Cardoso:2005vk,Dolan:2007mj}. For this purpose continued fraction method is used, which shows that $\omega_I$ takes the maximum value $\omega_I/M \sim 1.5 \times 10^{-7}$ at $l=m=1,a/M \sim1, \mu M \sim 0.42$. We confirmed that our new code using the continued fraction methods to evaluate $\omega_I$ consistently reproduces the results of Ref.~\cite{Dolan:2007mj}.

\section{Renormalization group analysis of an axion cloud}
\label{section:3}

In this section, we formulate how to apply the RG method to the time evolution of an axion cloud. We are interested in whether the dissipative effect of self-interaction terminates the instability or not. Therefore, we will develop our formulation to second order perturbation theory, where the dissipative effect first appears. We assume that one can neglect the gravitational perturbation caused by the axion field, for simplicity.

\subsection{Derivation of the Evolution equation}\label{sec:3.1}

To consider the effect of the axion self-interaction, we solve Eq.~\eqref{eom1} using perturbation theory. We assume that the amplitude of axion field is small enough, which allows us to approximate Eq.~\eqref{eom1} as
\begin{align}\label{eom2}
	(\square_g - \mu^2)\phi + \lambda \phi^3 = 0~,
\end{align}
where $\lambda = \mu^2/6 F_a^2$. This means that we have approximated the potential of the axion as
\begin{align}\label{potential}
    V(\phi) = \mu^2 F_a^2 \left(1 - \cos\left(\frac{\phi}{F_a}\right) \right)\sim \frac{\mu^2}{2}\phi^2 - \frac{\mu^2}{4! F_a^2}\phi^4~.
\end{align}
The negative sign of the term proportional to $\phi^4$ clearly indicates that the self-interaction of the axion is attractive.

We solve Eq.~\eqref{eom2} by expanding the solution in $\lambda$ as
\begin{align}\label{phiexpand}
	\phi = \phi_{(0)} +\lambda \phi_{(1)}+\lambda^2 \phi_{(2)}+\cdots~.
\end{align}
Substituting Eq.~\eqref{phiexpand} to Eq.~\eqref{eom2}, we obtain,
\begin{align}
	(\square_g - \mu^2)\phi_{(0)} &= 0~,\\
	\label{eomfirstorder}
	(\square_g - \mu^2)\phi_{(1)} &= - \phi_{(0)}^3~,\\
	\label{eomsecondorder}
	(\square_g - \mu^2)\phi_{(2)} &= - 3  \phi_{(0)}^2 \phi_{(1)}~.
\end{align}
Since we are interested in the dynamical evolution of an axion cloud, we take the fastest growing superradiant mode as the zeroth order solution:
\begin{align}\label{zerothphi}
	\phi_{(0)} = A(t_0)e^{-i(\omega_0 t - m_0 \varphi)}S_{l_0m_0\omega_0}(\theta)R_{l_0m_0\omega_0}(r) + \mathrm{c.c.}~,
\end{align}
where $S_{l_0m_0\omega_0}(\theta)$ and $R_{l_0m_0\omega_0}(r)$ are the solutions of Eqs.~\eqref{angularEOM} and \eqref{radialEOM} with $\omega_{0,I} > 0$. The label $(l_0,m_0,\omega_0)$ specifies the fastest growing superradiant mode that we are considering. $A(t_0)$ is the amplitude of the cloud at $t = t_0$, and c.c. denotes the complex conjugate. Here, we introduce an arbitrary reference time $t_0$, which is used in the RG method later. We first demonstrate the use of the RG method with the first order perturbation equation~\eqref{eomfirstorder} in Sec.~\ref{subfirst}, and then proceed to the second order in Sec.~\ref{subsecond}.

\subsubsection{First order perturbation}\label{subfirst}

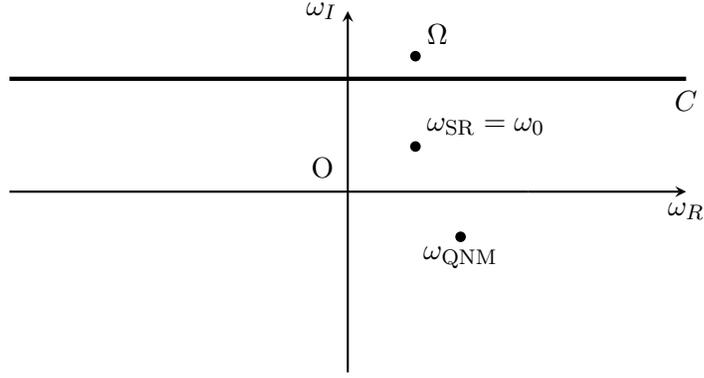
\begin{figure}
\centering
	\begin{tikzpicture}[x=3cm,y=3cm]
	 \coordinate (C) at (0.5,-0.5) node at (C) [above left=0.5mm] {O};
	 \coordinate (O) at (1.0,-0.7) node at (O) [below] {$\omega_{\mathrm{QNM}}$};
	 \coordinate (A) at (1.3,-0.5) node at (A) [below=0] {};
	 \coordinate (B) at  (0.8,-0.3) node at (B) [above right=0.2mm] {$\omega_{\mathrm{SR}} = \omega_0$};
	 \coordinate (S1) at  (0.8,0.1) node at (S1) [above right=0.2mm] {$\Omega$};
	
	\draw [ultra thick](-1.0,0.0)--(2.0,0.0) node[anchor=north]{$C$};
	 \draw [thick](-1.0,-0.5)--(A);
	 \draw [thick, -stealth] (C) --(0.5,0.3) node [anchor=east]{$\omega_I$};
 \draw[thick,-stealth] (A) -- (2.0,-0.5)
 node[anchor=north]{$\omega_R$};
\draw[thick] (C) -- (0.5,-1.3); 
10
     \fill (O) circle [radius=2pt];
     \fill (B) circle [radius=2pt];
 \fill (S1) circle [radius=2pt];
\end{tikzpicture}
\caption{Line $C$ shows the contour of integration on the complex plane of the frequency $\omega$. $\omega_{\rm SR}$ and $\omega_{\rm QNM}$ denote the frequency of the superradiant mode and that of the quasi-normal mode, respectively. The frequency $\Omega$ denotes the frequency of the source.}
\label{fig:1}
\end{figure}

Using the retarded Green's function $G_{\mathrm{ret}}$, the first order equation \eqref{eomfirstorder} is formally solved as
\begin{align}
	\label{firstpertgreen}
	\phi_{(1)} = - \int\sqrt{-g(x')}\, d^4 x' \,  G_{\mathrm{ret}}(x,x') \phi_{(0)}^3(x') + \mathrm{(initial\ value)}~.
\end{align}
Here, the Green's function is defined to satisfy
\begin{align}
	 (\square_g - \mu^2)G_{\mathrm{ret}}(x,x') = \frac{1}{\sqrt{-g(x)}}\delta^{(4)}(x-x')~.
\end{align}
Separation of variables on Kerr space-time allows spectral decomposition of the Green's function as 
\begin{gather}\label{specdecomp}
		G_\mathrm{ret}(x,x') = \frac{1}{2\pi}\sum_{l,m}\int_C\frac{d\omega}{2\pi}e^{-i\omega(t-t')}e^{im(\varphi-\varphi')}S_{lm\omega}(\theta)S_{lm\omega}(\theta')G^\omega_{lm}(r,r')~,
\end{gather}
where
\begin{align}
	\label{modegreen}
	G^\omega_{lm}(r,r') = \frac{1}{W_{lm}(\omega)}\left(R^{r_+}_{lm\omega}(r)R^{\infty+}_{lm\omega}(r')\theta(r'-r) +R^{r_+}_{lm\omega}(r')R^{\infty+}_{lm\omega}(r)\theta(r-r')  \right)\,,
\end{align}
and the function $W_{lm}(\omega)$ is the Wronskian of $R^{r_+}$ and $R^{\infty+}$ defined by
\begin{align}
	W_{lm}(\omega) = \Delta\left(R^{r_+}\partial_r R^{\infty+} - R^{\infty+}\partial_r R^{r_+}\right)~.
\end{align}
The integration contour $C$ is as shown in Fig.~\ref{fig:1}. As the integral over $\omega$ in Eq.~\eqref{specdecomp} picks up poles of the integrand, which correspond to zeros of the Wronskian, we need to choose the integration contour $C$ to pass above all the poles so that the Green's function satisfies the retarded boundary condition. 

We introduced $R^{r_+}$ and $R^{\infty\pm}$ as the solutions of Eq.~\eqref{radialEOM} satisfying the boundary conditions
\begin{subequations}
\label{BC}
\begin{align}
	\label{inhorizon}
	R^{r_+} &\longrightarrow \begin{cases}
	e^{-i (\omega-m\Omega_H)r_*}~, & (r\to r_+)\\
	A_{\mathrm{in}}(\omega) \frac{e^{-i \sqrt{\omega^2-\mu^2}r_*}}{r} + A_{\mathrm{out}}(\omega)\frac{e^{+i\sqrt{\omega^2-\mu^2}r_*}}{r}~, & (r\to+\infty)
	\end{cases}~,\\
	\label{outinf}
	R^{\infty+} &\longrightarrow \begin{cases}
	B_{\mathrm{in}}(\omega)e^{-i (\omega-m\Omega_H)r_*} + B_{\mathrm{out}}(\omega)e^{+i (\omega-m\Omega_H)r_*}~, & (r\to r_+)\\
	\frac{e^{+i\sqrt{\omega^2-\mu^2}r_*}}{r}~, & (r\to+\infty)
	\end{cases}~,\\
	\label{ininf}
	R^{\infty-} &\longrightarrow \begin{cases}
	B^*_{\mathrm{out}}(-\omega^*)e^{-i (\omega-m\Omega_H)r_*} + B^*_{\mathrm{in}}(-\omega^*)e^{+i (\omega-m\Omega_H)r_*}~, & (r\to r_+)\\
	\frac{e^{-i\sqrt{\omega^2-\mu^2}r_*}}{r}~, & (r\to+\infty)
	\end{cases}~,
\end{align}
\end{subequations}
where the superscript ``$^*$" denotes the complex conjugate of the variable and $r_*$ is the tortoise coordinate defined by $dr_* =  (r^2 + a^2)\Delta^{-1} dr$. The boundary condition for $R^{r_+}$ (see Eq.~\eqref{inhorizon}) implies that there exist only ingoing waves at the outer horizon and that for $R^{\infty \pm}$ (see Eq.~\eqref{outinf}) means purely outgoing/ingoing waves at infinity. 
The asymptotic forms at $r\to\infty$ (see Eqs.~\eqref{inhorizon} and \eqref{outinf}) determine the value of Wronskian as
\begin{align}
	W_{lm}(\omega) = 2 i \sqrt{\omega^2-\mu^2}A_{in}(\omega)~.
\end{align}
At the frequency of the superradiant mode $\omega = \omega_0$, $W_{lm}(\omega)$ vanishes. This is because superradiant modes satisfy ingoing boundary condition at the outer horizon and decaying boundary condition at infinity (we chose $\Im{\sqrt{\omega^2 - \mu^2}}>0$ branch). 

Substituting Eq.~\eqref{specdecomp} into Eq.~\eqref{firstpertgreen}, we obtain
\begin{align}\label{firstordspect}
\baligned{}
	\phi_{(1)} &=  -\frac{1}{2\pi}\sum_{l,m}\int_{0}^{2\pi} d\varphi' \int_{-1}^{1}d\cos\theta'\int_{r_+}^{+\infty}dr' \ (r'^2 + a^2 \cos^2\theta')  \cr
	& \ \ \ \ \ \times\int_{-\infty}^{t}dt'\int_C\frac{d\omega}{2\pi}e^{-i\omega(t-t')}e^{im(\varphi-\varphi')}S_{lm\omega}(\theta)S_{lm\omega}(\theta')G^\omega_{lm}(r,r')\phi_{(0)}^3(x')~.
\ealigned{}
\end{align}
Integration over $\varphi'$ gives $\delta_{mm'}$. In the present case, the source $\phi_{(0)}^3$ (see Eq.~\eqref{zerothphi} for the definition of $\phi_{(0)}$) contains only $m = m_0$ and $m = 3m_0$ components. Therefore, the summation over $m$ only picks up $m=m_0$ and $m=3 m_0$. Notice that if we take the integration contours carefully so that both $t'$ and $\omega$ integrals converge, we can exchange their order. We first perform the integration with respect to $t'$. When $\Im{\Omega - \omega} > 0$, which is satisfied by taking the contour $C$ as shown in Fig.~\ref{fig:1}, the integration over $t'$ gives
\begin{align}
	e^{-i\omega t}\int^t_{-\infty} dt'\,e^{i(\omega - \Omega)t'} = \frac{e^{-i\Omega t}}{i(\omega -\Omega)}~.
\end{align}
Next, we integrate over $\omega$. The form of this integration is
\begin{align}
	\int_C \frac{d\omega}{2\pi i}\frac{e^{-i\Omega t}}{\omega -\Omega} S_{lm\omega}(\theta)S_{lm\omega}(\theta')G^\omega_{lm}(r,r')\phi_{(0)}^3(x')~.
\end{align}
At large $|\omega|$, the integrand falls faster than $|\omega|^{-2}$, if the mode functions $R_{lm\omega}$ and $S_{lm\omega}$ do not grow faster than $|\omega|^0$ as $|\omega| \to \infty$. This $|\omega|^{-2}$ factor partly comes from $1/\omega$ factor in the Wronskian.

Now, we modify the integration contour $C$ to a large semi-circle in the upper half plane and a small circle around the pole at $\omega = \Omega$. As we have stated above, the integrand falls off as fast as $\propto |\omega|^{-2}$, the integral at a large semicircle vanishes. Therefore, all contributions to the integral come from the pole at $\omega = \Omega$. 

In Eq.~\eqref{firstordspect} the source contains frequencies with $\omega = \omega_0+2i\omega_{0,I}$ and $3\omega_0$, which are the only frequencies that appear in the first order particular solution:
\begin{align}
\label{firstpertnonren}
\baligned{}
	\phi_{(1)}(x) = - \sum_l &\left[A^3 e^{-3i(\omega_0 t - m_0 \varphi)}S_{l3m_03\omega_0}(\theta) f^{(1)}_{l3m_03\omega_0}(r) \right.\cr
	& \left.+ 3 A|A|^2 e^{2\omega_{0,I}t}e^{-i(\omega_0 t - m_0 \varphi)}S_{lm_0\omega_0+2i\omega_{0,I}}(\theta)f^{(1)}_{lm_0\omega_0+2i\omega_{0,I}}(r) \right]+ \mathrm{c.c.}\cr
	& + (\mathrm{homogeneous\ solution})~.
\ealigned{}
\end{align}
Here, $f^{(1)}_{l3m_03\omega_0}(r)$ and $f^{(1)}_{lm_0\omega_0+2i\omega_{0,I}}(r)$ are defined by 
\begin{align}
\baligned{}
f^{(1)}_{l3m_0\omega}(r) \equiv &\int dr'd\cos\theta'\ (r'^2 + a^2 \cos^2\theta')S_{l3m_0\omega}(\theta')G^\omega_{l3m_0}(r,r')\cr
&\times S_{l_0m_0\omega_0}(\theta')^3 R_{l_0m_0\omega_0}(r')^3~,
\ealigned{}\\
\baligned{}
\label{flm0}
f^{(1)}_{lm_0\omega}(r) \equiv& 
\int dr'd\cos\theta'(r'^2 + a^2 \cos^2\theta')S_{lm_0\omega}(\theta')G^\omega_{lm_0}(r,r')\cr
 & \times |S_{l_0m_0\omega_0}(\theta')|^2 S_{l_0m_0\omega_0}(\theta')|R_{l_0m_0\omega_0}(r')|^2 R_{l_0m_0\omega_0}(r')~.
 \ealigned{}
\end{align}
It is easy to see from Eq.~\eqref{firstpertnonren} that the perturbative solution grows exponentially so that the solution would break the assumption that the amplitude of the perturbation is small. We use the RG method to avoid the breakdown of the perturbative expansion (see appendix \ref{App:A} for review on the RG method).

To apply the RG method, we identify the term that diverges in $\omega_{0,I}\to0$ limit. Note that we are considering the superradiant mode which satisfies $\omega_{0,I}/\omega_{0,R} \ll 1$. Recall that the Wronskian $W_{lm}(\omega)$ contained in the Green's function \eqref{specdecomp} has a zero at $(\omega, l, m) = (\omega_0, l_0, m_0)$. From the expression \eqref{flm0}, we see that the Green's function in $f^{(1)}_{l_0m_0\omega_0 + 2i\omega_{0,I}}(r)$ contains the factor $1/W_{l_0m_0}(\omega_0 + 2i\omega_{0,I})$, which diverges for $\omega_{0,I}\to 0$. Taking the leading term in the Taylor expansion of the Wronskian around $\omega=\omega_0$,
\begin{align}
	\label{wronskianalphadef}
	W_{lm}(\omega)\sim2i\sqrt{\omega_0^2 -\mu^2}\alpha_{\omega_0}(\omega-\omega_0) + \dots~,
\end{align}
we can extract the leading term in $f^{(1)}_{l_0m_0\omega_0 + 2i\omega_{0,I}}(r)$ in the limit $\omega_{0,I} \to 0$ as 
\begin{align}
\label{f1ord1}
\baligned{}
	f^{(1)}_{l_0m_0\omega_0+2i\omega_{0,I}}(r) \sim & \int dr' d\cos\theta' (r'^2 + a^2 \cos^2\theta') |S_{l_0m_0\omega_0}(\theta')|^2 S_{l_0m_0\omega_0}(\theta')^2\cr
	&\ \ \times\frac{R_{l_0m_0\omega_0}(r) R_{l_0m_0\omega_0}(r')}{2i\alpha_{\omega_0}\sqrt{\omega_0^2 - \mu^2}2 i \omega_{0,I}A_{\mathrm{out}}}|R_{l_0m_0\omega_0}(r')|^2 R_{l_0m_0\omega_0}(r')\cr
	\equiv&\, C^{(1)}_{l_0m_0\omega_0} R_{l_0m_0\omega_0}(r)~,
\ealigned{}
\end{align}
which we call a divergent term, although, strictly speaking, it is not divergent since $\omega_{0,I}$ is small but not vanishing.

We can claim that the term of $\mathcal{O}\left(\omega_{0,I}^{-1}\right)$ in  $C^{(1)}_{l_0m_0\omega_0}$ is purely imaginary. In the limit $\omega_{0,I} \to 0$, the equations for the radial and angular mode functions \eqref{radialEOM} and \eqref{angularEOM} are both real, and hence $R_{l_0m_0\omega_0}$ and $S_{l_0m_0\omega_0}$ can be chosen to be real. Also,  $\alpha_{\omega_0}$ and $A_{\mathrm{out}}$, which are calculated through the mode functions, are real in this limit.  Thus, the leading term in $C^{(1)}_{l_0m_0\omega_0}$ is real except for the factor $\sqrt{\omega_0^2 - \mu^2}$, which is purely imaginary in the limit, $\omega_{0,I}\to 0$, because a supperradiant state is a bound state satisfying $\omega_{0,R} < \mu$. To summarize, we have
\begin{align}
    \Re{C^{(1)}_{l_0m_0\omega_0}} &= \mathcal{O}(\omega_{0,I}^0)~,\\
    \Im{C^{(1)}_{l_0m_0\omega_0}} &= \mathcal{O}(\omega_{0,I}^{-1})~.
\end{align}

Using the RG method, we cancel this divergence of $\Im{C^{(1)}_{l_0m_0\omega_0}}$ by adding the homogeneous solution with an appropriate amplitude. This procedure is the same as inserting the counterterms in the renormalization in quantum field theory (QFT). After this procedure, the first order perturbative solution is given by
\begin{align}
\label{firstordersolnonren}
\baligned{}
	\phi =& \left(A +3\lambda \tilde{C}^{(1)}_{l_0m_0\omega_0} A |A|^2 e^{2\omega_{0,I}t_0}\right)e^{-i(\omega_0 t - m_0 \varphi)}S_{l_0m_0\omega_0}(\theta) R_{l_0m_0\omega_0}(r)\cr
	&\ \ \ -\lambda \left(\sum_{l=3m_0}^{\infty} \left[A^3 e^{-3i(\omega_0 t - m_0 \varphi)}S_{l3m_03\omega_0}(\theta) f^{(1)}_{l3m_03\omega_0}(r) \right]\right.\cr
	& \ \ \  \ \ \left.+ \sum_{l=m_0}^\infty \left[3 A|A|^2 e^{2\omega_{0,I}t}e^{-i(\omega_0 t - m_0 \varphi)}S_{lm_0\omega_0+2i\omega_{0,I}}(\theta)f^{(1)}_{lm_0\omega_0+2i\omega_{0,I}}(r) \right] \right)\cr
	& \ \ \ \ \ \ \ + \mathrm{c.c.}~,
\ealigned{}
\end{align}
where
\begin{align}
    \tilde{C}^{(1)}_{l_0m_0\omega_0} \equiv C^{(1)}_{l_0m_0\omega_0} + \delta C^{(1)}_{l_0m_0\omega_0}~.
\end{align}
We explicitly added $\delta C^{(1)} (= \mathcal{O}(\omega_{0,I}^0))$ to express the arbitrariness in identifying the non-divergent part (see App.\ref{App:B} for the relation between the choice of $\delta C^{(1)}$ and the definition of the amplitude $A(t)$). This is the same as choosing the scheme of renormalization in QFT. 

Now, we demand that the expression \eqref{firstordersolnonren} satisfies the RG equation:
\begin{align}
	\PDD{\phi}{t_0} = 0~.
\end{align}
We substitute Eq.~\eqref{firstordersolnonren} into the RG equation, to obtain the evolution equation for the amplitude as
\begin{align}
	\ODD{A(t_0)}{t_0} = -6\lambda \omega_{0,I} \tilde{C}^{(1)}_{l_0m_0\omega_0}A(t_0) |A(t_0)|^2 e^{2\omega_{0,I}t_0} + \mathcal{O}(\lambda^2)~.
\end{align}
Redefining the amplitude to incorporate the exponentially growing factor contained in the zeroth order solution $\phi_{(0)}$ will be more convenient. Namely, we define $\mathcal{A}(t)\equiv A(t)e^{\omega_{0,I} t}$, and then $\mathcal{A}$ satisfies the following equation,
\begin{align}\label{eq:41}
	\ODD{\mathcal{A}}{t} = \omega_{0,I}\mathcal{A}-6\lambda \omega_{0,I} \tilde{C}^{(1)}_{l_0m_0\omega_0}\mathcal{A} |\mathcal{A}|^2 ~.
\end{align}
We decompose this equation into the amplitude and phase part by writing $\mathcal{A}$ in the form $\mathcal{A} = |\mathcal{A}|(t) e^{- i \Theta(t)}$ to obtain
\begin{align}\label{amp1st}
	\ODD{|\mathcal{A}|}{t} &= \omega_{0,I}|\mathcal{A}| - 6 \lambda\omega_{0,I}\Re{\tilde{C}^{(1)}_{l_0m_0\omega_0}}  |\mathcal{A}|^3~,\\
	\label{phase1st}
	\ODD{\Theta}{t} &=  6 \lambda \omega_{0,I} \Im{\tilde{C}^{(1)}_{l_0m_0\omega_0}} |\mathcal{A}|^2~.
\end{align}

Substituting the solution of Eq.~\eqref{eq:41} to Eq.~\eqref{firstordersolnonren}, and setting $t_0 = t$, we finally obtain the renormalized first order perturbative solution as
\begin{align}
\label{renomfirstorder}
\baligned{}
	\phi =& \mathcal{A}(t) e^{-i(\omega_{0,R} t - m_0 \varphi)}S_{l_0m_0\omega_0}(\theta) R_{l_0m_0\omega_0}(r)\cr
	&\ \ \ -\lambda \left(\sum_{l=3m_0}^{\infty} \mathcal{A}^3(t) e^{-3i(\omega_{0,R} t - m_0 \varphi)}S_{l3m_03\omega_0}(\theta) f^{(1)}_{l3m_03\omega_0}(r) \right.\cr
	& \ \ \  \ \ \left.+ \sum_{l > m_0}^{\infty} \left[3 \mathcal{A}(t)|\mathcal{A}(t)|^2 e^{-i(\omega_{0,R} t - m_0 \varphi)}S_{lm_0\omega_0+2i\omega_{0,I}}(\theta)f^{(1)}_{lm_0\omega_0+2i\omega_{0,I}}(r)\right]\right.\cr
	&\ \ \ \ \ \ \ + 3 \mathcal{A}(t)|\mathcal{A}(t)|^2 e^{-i(\omega_{0,R} t - m_0 \varphi)}\delta\phi_{(1)}\Biggr) +\mathrm{c.c.}~,
\ealigned{}
\end{align}
where 
\begin{align}\label{delphi1}
\baligned{}
	\delta\phi_{(1)} &\equiv S_{l_0m_0\omega_0+2i\omega_{0,I}}(\theta)f^{(1)}_{l_0m_0\omega_0+2i\omega_{0,I}}(r) - \tilde{C}^{(1)}_{l_0m_0\omega_0}S_{l_0m_0\omega_0}(\theta)R_{l_0m_0\omega_0}(r)
\ealigned{}
\end{align}
is the non-divergent part of $f^{(1)}_{l_0m_0\omega_0 + 2i\omega_{0,I}}(r)$. Two types of modes appear in Eq.~\eqref{renomfirstorder}. The first term in the parentheses has the frequency $3 \omega_{0,R} > \mu$, which means that the mode is unbounded and can dissipate to infinity. The second and third terms have the frequency $\omega_{0,R} < \mu$, and thus these are bounded and cannot dissipate energy to infinity. Also, an important property common to all modes is that they satisfy superradiant condition $\omega < m \Omega_H$. Therefore, dissipating the energy of the cloud back to the black hole does not occur in our setup describing the adiabatic evolution starting with a single dominant superradiant mode.

\subsubsection{Second order perturbation}\label{subsecond}

Now, we proceed to the second order analysis to incorporate the dissipative effect from scattering via self-interaction. This calculation is almost in parallel to the first order one: we first solve Eq.~\eqref{eomsecondorder} formally, then identify the divergent part in the formal solution, and finally apply the RG method to eliminate this divergence.

A formal solution of Eq.~\eqref{eomsecondorder} is given by
\begin{align}
	\phi_{(2)} = -3 \int\sqrt{-g(x')} d^4 x' G_{\mathrm{ret}}(x,x') \phi_{(0)}^2\phi_{(1)}  + \mathrm{(initial\ value)}~.
\end{align}
As we have seen before, the part that diverges in the limit $\omega_{0,I} \to 0$ originates from the Wronskian $W_{l_0m_0}(\omega_0)$. Using the spectral representation \eqref{specdecomp}, we see that $\phi_{(2)}$ contains the Wronskian in the following part of the solution:
\begin{align}
\baligned{}
	\phi_{(2)} \supset& 3\sum_{l,l'}\int d\cos\theta' \int dr' (r'^2+a^2 \cos^2\theta') A|A|^4 e^{4\omega_{0,I}t}e^{-i(\omega_0 t - m_0\varphi)}\cr
	&\times S_{lm_0\omega_0+4i\omega_{0,I}}(\theta)S_{lm_0\omega_0+4i\omega_{0,I}}(\theta') G^{\omega_0+4i\omega_{0,I}}_{lm_0}(r,r')\cr
	&\ \ \ \times \left(S_{l_0m_0\omega_0}^*(\theta')^2R_{l_0m_0\omega_0}^*(r')^2S_{l'3m_03\omega_0}(\theta')f^{(1)}_{l'3m_03\omega_0}(r') \right.\cr
	&\ \ \ \ \ \left.+ 6 |S_{l_0m_0\omega_0}(\theta')|^2|R_{l_0m_0\omega_0}(r')|^2 S_{l'm_0\omega_0+2i\omega_{0,I}}(\theta')f^{(1)}_{l'm_0\omega_0+2i\omega_{0,I}}(r')\right.\cr
	&\ \ \ \ \ \ \  + \left. 3S_{l_0m_0\omega_0}(\theta')^2 R_{l_0m_0\omega_0}(r')^2 S^*_{l'm_0\omega_0+2i\omega_{0,I}}(\theta')f^{(1)*}_{l'm_0\omega_0+2i\omega_{0,I}}(r')\right)~.
\ealigned{}
\end{align}
From this expression we identify the divergent part as
\begin{align}
	3 C^{(2)}_{l_0m_0\omega_0}A|A|^4e^{4\omega_{0,I}t}e^{-i(\omega_0 t - m_0 \varphi)}S_{l_0m_0\omega_0} R_{l_0m_0\omega_0}~,
\end{align}
where $C^{(2)}$ is defined by
\begin{align}\label{defC2}
	\baligned{}
	C^{(2)}_{l_0m_0\omega_0} = &\frac{1}{2i\alpha_{\omega_0}\sqrt{\omega_0^2 - \mu^2} 4i\omega_{0,I}A_{out}}\sum_{l'}\int d\cos\theta'\int dr'\cr
	&\times (r'^2 + a^2 \cos^2\theta')S_{l_0m_0\omega_0}(\theta') R_{l_0m_0\omega_0}(r')\cr
	&\times\left(S_{l_0m_0\omega_0}^*(\theta')^2R_{l_0m_0\omega_0}^*(r')^2S_{l'3m_03\omega_0}(\theta')f^{(1)}_{l'3m_03\omega_0}(r') \right.\cr
	&\ \ \ \left.+ 6 |S_{l_0m_0\omega_0}(\theta')|^2|R_{l_0m_0\omega_0}(r')|^2 S_{l'm_0\omega_0+2i\omega_{0,I}}(\theta')f^{(1)}_{l'm_0\omega_0+2i\omega_{0,I}}(r')\right.\cr
	&\ \ \ \ \ \ \left.+ 3S_{l_0m_0\omega_0}(\theta')^2 R_{l_0m_0\omega_0}(r')^2 S^*_{l'm_0\omega_0+2i\omega_{0,I}}(\theta')f^{(1)*}_{l'm_0\omega_0+2i\omega_{0,I}}(r')\right)~.
	\ealigned{}
\end{align}
Same as first order solution (see Eq.~\eqref{renomfirstorder} and discussion below), there are two contributions to $C^{(2)}$. One is due to $\omega > \mu$ modes, which dissipate energy to infinity and the other is due to $\omega < \mu$ modes, which decay at infinity and conserve energy. 

There is another type of source which produces divergence in the second order solution.  This source comes from the homogeneous solution that we added to eliminate the divergent part in the first order solution. Using the Green's function, the contribution to the second order solution from this type of source is 
\begin{align}
\baligned{}
	-9 |A|^2 &e^{2\omega_{0,I}t_0} \int\sqrt{-g(x')}\, d^4 x' G_{\mathrm{ret}}(x,x')\,\phi_{(0)}^2(x') \cr
	& \times\left( A \tilde{C}^{(1)}_{l_0m_0\omega_0}e^{-i(\omega_0 t' - m_0 \varphi')}S_{l_0m_0\omega_0}(\theta')R_{l_0m_0\omega_0}(r') + \mathrm{c.c.}\right) ~.
\ealigned{}
\end{align}
We identify the divergent part in this expression as
\begin{align}
\baligned{}
	-9 A|A|^4 e^{2\omega_{0,I}t_0}e^{2\omega_{0,I}t}\left(2 C^{(1)}_{l_0m_0\omega_0}\tilde{C}^{(1)}_{l_0m_0\omega_0}+\tilde{C}^{(1)*}_{l_0m_0\omega_0}C^{(1)}_{l_0m_0\omega_0}\right) \cr
	\times e^{-i(\omega_{0} t - m_0 \varphi)}S_{l_0m_0\omega_0}(\theta) R_{l_0m_0\omega_0}(r) 
	&+ \mathrm{c.c.}~.
\ealigned{}
\end{align}

We choose the initial condition for $\phi_{(2)}$ at $t = t_0$ to eliminate the divergence derived above. This turns out to be adding the homogeneous solution
\begin{align}
\label{renomamp}
\baligned{}
	- 3 \lambda^2 \left(\left(C^{(2)}_{l_0m_0\omega_0}+ \delta C^{(2)}_{l_0m_0\omega_0}\right) \right.&\left.- 3\left(2 C^{(1)}_{l_0m_0\omega_0}\tilde{C}^{(1)}_{l_0m_0\omega_0}+\tilde{C}^{(1)*}_{l_0m_0\omega_0}C^{(1)}_{l_0m_0\omega_0}\right)\right) \cr
	&\times A|A|^4 e^{4\omega_{0,I}t_0} e^{-i(\omega_0 t - m_0 \varphi)}S_{l_0m_0\omega_0} R_{l_0m_0\omega_0}~,
\ealigned{}
\end{align}
to $\phi_{(2)}$. Here, $\delta C^{(2)}_{l_0m_0\omega_0}$ represents the arbitrariness in the choice of the non-divergent part. 

Now, we can derive the amplitude equation by imposing the RG equation as in the first order case. Taking care of the time dependence of the amplitude, we obtain
\begin{align}
	\label{secondorderren}
\baligned{}
	\ODD{A}{t} + &\, 3 \lambda \delta C^{(1)}_{l_0m_0\omega_0} \left(2 |A|^2\ODD{A}{t} + A^2 \ODD{A^*}{t}\right)e^{2\omega_{0,I}t} 
\cr	&= - 6 \lambda \omega_{0,I} \tilde{C}_{l_0m_0\omega_0}^{(1)}A|A|^2 e^{2\omega_{0,I}t} 
	 + 12\lambda^2 \omega_{0,I} \tilde{C}_{l_0m_0\omega_0}^{(2)} A|A|^4 e^{4\omega_{0,I}t}~,
\ealigned{}
\end{align}
where $\tilde{C}^{(2)}_{l_0m_0\omega_0}$ is defined by
\begin{align}\label{eq:58}
	\tilde{C}^{(2)}_{l_0m_0\omega_0} &\equiv \hat{C}^{(2)}_{l_0m_0\omega_0} + \delta C^{(2)}_{l_0m_0\omega_0}-  \frac{3}{2} C^{(1)}_{l_0m_0\omega_0} \left(2 \delta C^{(1)}_{l_0m_0\omega_0}+\delta C^{(1)*}_{l_0m_0\omega_0}\right) ~,
\end{align}
and $\hat{C}^{(2)}_{l_0m_0\omega_0}$ is defined by
\begin{align}
	\hat{C}^{(2)}_{l_0m_0\omega_0} &\equiv C^{(2)}_{l_0m_0\omega_0}-  \frac{3}{2} C^{(1)}_{l_0m_0\omega_0} \left(2 C^{(1)}_{l_0m_0\omega_0}+C^{(1)*}_{l_0m_0\omega_0}\right) ~.
\end{align}
From the expression~\eqref{eq:58}, it seems that the divergence in the second order perturbation could be eliminated by adjusting the first order counterterm $\delta C^{(1)}_{l_0m_0\omega_0}$. But we will see in appendix \ref{App:B} that this is not true, at least for the real part of the divergence.

We see the cancellation of terms of $\mathcal{O}(\omega_{0,I}^{-2})$ in $\hat{C}^{(2)}_{l_0m_0\omega_0}$. In Eq.~\eqref{defC2}, the $\mathcal{O}(\omega_{0,I}^{-2})$ terms come from $l=l_0,m=m_0$ mode, which are 
\begin{align}
\baligned{}
	C_{l_0m_0\omega_0}^{(2)}\bigl|_{\mathcal{O}(1/\omega_I^2)}=&\frac{1}{2i\alpha_{\omega_0}\sqrt{\omega_0^2 - \mu^2} 4i\omega_{0,I}A_{out}}\int d\cos\theta'\int dr' \cr
	&\times (r'^2 + a^2 \cos^2\theta')S_{l_0m_0\omega_0}(\theta') R_{l_0m_0\omega_0}(r')\cr
	&\times\left(6 |S_{l_0m_0\omega_0}(\theta')|^2|R_{l_0m_0\omega_0}(r')|^2 S_{l_0m_0\omega_0+2i\omega_{0,I}}(\theta')f^{(1)}_{l_0m_0\omega_0+2i\omega_{0,I}}(r')\right.\cr
	&\ \ \ \left.+ 3S_{l_0m_0\omega_0}(\theta')^2 R_{l_0m_0\omega_0}(r')^2 S^*_{l_0m_0\omega_0+2i\omega_{0,I}}(\theta')f^{(1)*}_{l_0m_0\omega_0+2i\omega_{0,I}}(r')\right)~.
	\ealigned{}
\end{align}
Notice that $l=l_0,m=m_0$ contributions to $C^{(2)}_{l_0m_0\omega_0}$ and $\left(C^{(1)}_{l_0m_0\omega_0}\right)^2$ can be put together in the following form:
\begin{align}
\baligned{}
	&\frac{6}{2i\alpha_{\omega_0}\sqrt{\omega_0^2 - \mu^2} 4i\omega_{0,I}A_{out}}\int d\cos\theta'\int dr' (r'^2 + a^2 \cos^2\theta')S_{l_0m_0\omega_0}(\theta') R_{l_0m_0\omega_0}(r')\cr
	&\qquad \times |S_{l_0m_0\omega_0}(\theta')|^2|R_{l_0m_0\omega_0}(r')|^2 S_{l_0m_0\omega_0+2i\omega_{0,I}}(\theta')f^{(1)}_{l_0m_0\omega_0+2i\omega_{0,I}}(r')-  3 C^{(1)2}_{l_0m_0\omega_0}\cr
	&\quad = \frac{6}{2i\alpha_{\omega_0}\sqrt{\omega_0^2 - \mu^2} 4i\omega_{0,I}A_{out}}\int d\cos\theta'\int dr' (r'^2 + a^2 \cos^2\theta')\cr
	&\qquad \times S_{l_0m_0\omega_0}(\theta') |S_{l_0m_0\omega_0}(\theta')|^2R_{l_0m_0\omega_0}(r')|R_{l_0m_0\omega_0}(r')|^2\cr
	&\qquad\quad \times \left(S_{l_0m_0\omega_0+2i\omega_{0,I}}(\theta')f^{(1)}_{l_0m_0\omega_0+2i\omega_{0,I}}(r') - C^{(1)}_{l_0m_0\omega_0} S_{l_0m_0\omega_0}R_{l_0m_0\omega_0}\right)\cr
	&\quad = \frac{6}{2i\alpha_{\omega_0}\sqrt{\omega_0^2 - \mu^2} 4i\omega_{0,I}A_{out}}\int d\cos\theta'\int dr' (r'^2 + a^2 \cos^2\theta')\cr
	&\qquad \times S_{l_0m_0\omega_0}(\theta') |S_{l_0m_0\omega_0}(\theta')|^2R_{l_0m_0\omega_0}(r')|R_{l_0m_0\omega_0}(r')|^2\delta\phi_{(1)}~.
\ealigned{}
\end{align}
Because the function $\delta \phi_{(1)}$ does not contain $\mathcal{O}(\omega_{0,I}^{-1})$ factor (see Eq.~\eqref{delphi1}), we see that $\mathcal{O}(\omega_{0,I}^{-2})$ factor in $C^{(2)}_{l_0m_0\omega_0}$ is safely cancelled by $\left(C^{(1)}_{l_0m_0\omega_0}\right)^2$. 

Solutions of Eq.~\eqref{secondorderren} describe the long time behavior of the axion cloud, by which we can examine whether the self-interaction of the axion terminates the superradiant instability or not. To solve Eq.~\eqref{secondorderren}, we need the values of the coefficients $C^{(1)}_{l_0m_0\omega_0}$ and $C^{(2)}_{l_0m_0\omega_0}$. Unfortunately, we cannot calculate $C^{(1)}_{l_0m_0\omega_0}$ and $C^{(2)}_{l_0m_0\omega_0}$ analytically and must calculate them numerically, because of the complexity of the Kerr space-time. We present the results of numerical calculations in Sec.~\ref{section:4}. Note that calculations of these quantities must be accurate relatively up to $\mathcal{O}(\omega_{0,I}/\omega_{0,R})$ to capture the divergent part in the solution correctly.

\subsection{Behavior of the amplitude equation}\label{sec:3.1.3}

Before calculating the numerical coefficients, we see some analytic feature of the evolution equation \eqref{secondorderren}. In this section, we abbreviate the subscripts ${l_0, m_0, \omega_0}$, for brevity. First, we comment on the validity of the perturbative renormalization. In the same way as Eqs.~\eqref{amp1st} and \eqref{phase1st}, we can rewrite Eq.~\eqref{secondorderren} as the evolution equations of the amplitude and the phase by substituting $A = |\mathcal{A}|(t) e^{-\omega_{0,I} t - i \Theta(t)}$, to obtain
\begin{align}\label{amp2nda}
	\baligned{}
	&\frac{1 + 9 \lambda \Re{\delta C^{(1)}} |\mathcal{A}|^2}{\omega_{0,I}}\ODD{|\mathcal{A}|}{t} + 3\lambda  \Im{\delta C^{(1)}}|\mathcal{A}|^3 \frac{\delta \omega}{\omega_{0,I}} \cr
	&\qquad =|\mathcal{A}| - 6 \lambda \Re{\tilde{C}^{(1)}}  |\mathcal{A}|^3 + 9 \lambda \Re{\delta C^{(1)}} |\mathcal{A}|^3 + 12\lambda^2 \Re{\tilde{C}^{(2)}} |\mathcal{A}|^5 ~,
	\ealigned{}\\
	\label{phase2nda}
	\mbox{and}\cr
	&\baligned{}
	(1 + 3 \lambda \Re{\delta C^{(1)}} |\mathcal{A}|^2)\frac{\delta\omega}{\omega_{0,I}} - 9 \lambda \Im{\delta C^{(1)}}\frac{1}{\omega_{0,I}} |\mathcal{A}| \ODD{|\mathcal{A}|}{t} \cr
	&\qquad = 6 \lambda  \Im{\tilde{C}^{(1)}}|\mathcal{A}|^2-9 \lambda  \Im{\delta{C}^{(1)}}|\mathcal{A}|^2 - 12\lambda^2 \Im{\tilde{C}^{(2)}} |\mathcal{A}|^4\,,
	\ealigned{}
\end{align}
neglecting the terms higher order in $\lambda$. Here, the frequency shift $\delta \omega$ is defined by 
\begin{align}
	\delta \omega \equiv \ODD{\Theta}{t}~.
\end{align}
Explicit dependence on the ambiguous terms $\delta C^{(1)}$ and $\delta C^{(2)}$ up to $\mathcal{O}(\lambda^2)$ is
\begin{align}\label{eq:65}
	\baligned{}
	\frac{1}{\omega_{0,I}}\ODD{|\mathcal{A}|}{t} = &|\mathcal{A}|- 6 \lambda \Re{C^{(1)} + \delta C^{(1)}}  |\mathcal{A}|^3 + 12\lambda^2 \Re{\hat{C}^{(2)} + \delta C^{(2)}} |\mathcal{A}|^5\cr
	&  + 18 \lambda^2 \Re{\delta C^{(1)} \left(2 \delta C^{(1)} + \delta C^{(1)*}\right)}|\mathcal{A}|^5 ~,
	\ealigned{}\\
	\baligned{}\label{eq:66}
	\frac{\delta\omega}{ \omega_{0,I} } =  &6 \lambda\Im{C^{(1)} + \delta C^{(1)}}|\mathcal{A}|^2 - 12\lambda^2 \Im{\hat{C}^{(2)} + \delta C^{(2)}} |\mathcal{A}|^4\cr
	& + 18  \lambda^2 \Im{2 C^{(1)}\delta C^{(1)*}- \delta C^{(1)} \left(2 \delta C^{(1)} + \delta C^{(1)*}\right)}|\mathcal{A}|^4~.
	\ealigned{}
\end{align}
From Eqs.~\eqref{eq:65} and \eqref{eq:66}, we find that the evolution of the amplitude and the frequency shift depends on the renormalization scheme, {\it i.e.}, how to choose $\delta C^{(1)}$ and $\delta C^{(2)}$. In App.~\ref{App:B}, we show these choices are related to the definition of the amplitude. 

$\delta\omega$ is expected to have a scheme independent meaning. Therefore, the evolution equation for $\delta\omega$ 
\begin{align}
\baligned{}
	\frac{1}{\omega_{0,I}}\ODD{\delta \omega}{t} = &\, 2 \delta \omega \Biggl(1 - \left(\Im{\hat{C}^{(2)}  + \delta C^{(2)}  }+ 3 \Re{C^{(1)}  }\Im{C^{(1)}} \right.\cr
	&\qquad \left. + \Re{6 C^{(1)}   + 9 \delta C^{(1)}  }\Im{\delta C^{(1)}}\right) \frac{\delta \omega/\omega_{0,I}}{3\Im{\tilde{C}^{(1)}}^2}\Biggr)~,
\ealigned{}
\end{align}
becomes scheme independent except for the terms of $\mathcal{O}(\omega_{0,I})$. However, in order to obtain the evolution equation for $\delta\omega$ up to $\mathcal{O}(\lambda^2)$, we need to proceed our calculation to one order higher, 
which turned out to be technically quite challenging. 
Below, instead of pursuing to obtain the scheme independent equation valid up to $\mathcal{O}(\lambda^2)$, 
we choose some schemes by adding conditions that specify $\delta C^{(1)}$ and $\delta C^{(2)}$
without the knowledge of higher order perturbation, 
and see the qualitative behavior of the cloud.

\subsubsection{Minimal subtraction scheme}\label{sec:3.2.1}
One scheme is to choose $\delta C^{(1)} = 0$ and $\delta C^{(2)} = 0$, which corresponds to minimally subtracting the divergent terms. This choice makes renormalization group equation quite simple as one can see from Eqs.~\eqref{eq:65} and \eqref{eq:66}. After taking this scheme, the evolution equations of the amplitude and the phase are given by
\begin{align}\label{amp2nd}
	\frac{1}{\omega_{0,I}}\ODD{|\mathcal{A}|}{t} &= |\mathcal{A}| - 6 \lambda \Re{C^{(1)}}  |\mathcal{A}|^3 + 12\lambda^2 \Re{\hat C^{(2)}} 
	 |\mathcal{A}|^5~,\\
	\label{phase2nd}
	\frac{\delta\omega}{ \omega_{0,I} } &=  6 \lambda  \Im{C^{(1)}}|\mathcal{A}|^2 - 12\lambda^2 \Im{\hat C^{(2)}}|\mathcal{A}|^4~.
\end{align}
A qualitative behavior of the axion cloud can be read off from Eq.~\eqref{amp2nd}. When the amplitude is
\begin{align}\label{eq:61}
	|\mathcal{A}|^2 = \frac{\Re{C^{(1)}}\left(1 \pm \sqrt{1 - \frac{4 \Re{\hat{C}^{(2)}}}{3 \left|\Re{C^{(1)}}\right|^2}}\right)}{4 \lambda \Re{\hat C^{(2)}}}~,
\end{align}
the right hand side of Eq.~\eqref{amp2nd} vanishes. However, when Eq.~\eqref{eq:61} is not a positive real number, this immediately indicates the indefinite growth of the amplitude and eventual breakdown of the perturbative expansion. By contrast, if Eq.~\eqref{eq:61} is a positive real number, then evolution of the axion cloud may terminate at this amplitude. 

Let us analyze the case when the saturation occurs in detail. As we shall confirm later, the attractive nature of the self-interaction implies $\Re{C^{(1)}}<0$. This means that $\Re{C^{(1)}}/\lambda <0$ irrespectively of the signature of $\lambda$, which is assumed in the following discussion. Then, we can assume $\Re{\hat C^{(2)}} < 0$, since otherwise the right hand side of Eq.~\eqref{eq:61} is always negative.  
When $\Re{\hat C^{(2)}} \lesssim -\left|\Re{C^{(1)}}\right|^2$, the saturation occurs at 
\begin{align}\label{eq:saturation}
	|\mathcal{A}|^2 \sim \frac{1}{2 \lambda \sqrt{3 \left|\Re{\hat C^{(2)}}\right|}}~.
\end{align}
Here, we adopt the appropriate sign in the right hand side of Eq.~\eqref{eq:61} so as to make it positive.When $-\left|\Re{C^{(1)}}\right|^2\lesssim \Re{\hat{C}^{(2)}}<0 $, the saturation of amplitude 
occurs at 
\begin{align}
	|\mathcal{A}|^2 \sim \frac{1}{6\lambda |\Re{C^{(1)}}|}~.
\end{align}
The saturation is likely to occur in the weakly non-linear regime, if there is some hierarchy in the magnitudes of coefficients, {\it i.e.}, if $\left|\Re{\hat C^{(2)}}\right|\ll \left|\Re{{C}^{(1)}}\right|^2$ or $\left|\Re{\hat C^{(2)}}\right|\gg \left|\Re{{C}^{(1)}}\right|^2$. 
However, we should recall that we assumed $\Re{\hat C^{(2)}} < 0$ at the very beginning of this discussion, 
which we will find later not to be the case. 

\subsubsection{Dissipative scheme}\label{sec:3.2.2}
Other than the minimal subtraction scheme, we can take the scheme in which the time evolution of the amplitude is totally governed by the dissipative energy loss to the infinity. Instead of putting $\delta C^{(1,2)} = 0$, we demand that non-dissipative part in the right hand side of the amplitude equation \eqref{eq:65} is set to 0. To do so, we must identify the dissipative part of the $C^{(2)}$. With the aid of the flux-balance equation, we identify
\begin{align}\label{eq:C2diss}
    C^{(2)\diss}_{l_0m_0\omega_0} = \frac{\sqrt{9\omega_0^2 - \mu^2}\sum_{l}|j_{l3m_03\omega_0}|^2}{2 \pi (\omega_0 - m_0 \Omega_H) (r_+^2 + a^2) \left(\int d\theta |S_{l_0m_0\omega_0}|^2 \right)}~,
\end{align}
as the dissipative part in the $C^{(2)}$ (see appendix \ref{sec:AppC} for the detail of the derivation). We observe that $C^{(2)\diss}$ is real except for the imaginary part of $\omega_0$. Therefore, imaginary part is suppressed by $\omega_{0,I}$ and thus we treat $C^{(2)\diss}$ as a purely real number.

Now, the scheme which respects the dissipation is realized by taking
\begin{align}\label{eq:74}
    \Re{\delta C^{(1)}} &= -\Re{C^{(1)}}~,\\
    \Re{\delta C^{(2)}} &= -\Re{\hat{C}^{(2)\cons}} - \frac{9}{2}\Re{C^{(1)}}^2~.
\end{align}
Here, we defined the non-dissipative part of $\hat{C}^{(2)}$ as 
\begin{align}
    \hat{C}^{(2)\cons} \equiv \hat{C}^{(2)} - C^{(2)\diss}~.
\end{align}
Note that $\Re{C^{(1)}}$ is also non-dissipative, and suppressed by $\omega_{0,I}$ compared with $C^{(1)}$ itself. This fact explains why we choose $\Re{\delta C^{(1)}}$ so as to eliminate $\Re{C^{(1)}}$ in Eq.\eqref{eq:74}. Because $\Im{C^{(1)}}$ and $\Im{\hat{C}^{(2)}}$ start with  $\mathcal{O}(\omega_{0,I}^{-1})$, we cannot eliminate these by $\delta C^{(1)}$ and $\delta C^{(2)}$, which is $\mathcal{O}(\omega_{0,I}^0)$. Therefore, since the imaginary parts do not contribute to the evolution equation of the amplitude, we simply take 
\begin{align}
\Im{\delta C^{(1)}} = 0~,\\
\Im{\delta C^{(2)}} = 0~.
\end{align}

After taking this scheme, amplitude equations \eqref{eq:65} and \eqref{eq:66} are
\begin{align}
	\baligned{}\label{eq:ampdiss}
	\frac{1}{\omega_{0,I}}\ODD{|\mathcal{A}|}{t} &= |\mathcal{A}| + 12\lambda^2 C^{(2)\diss} |\mathcal{A}|^5 ~,
	\ealigned{}\\
	\baligned{}\label{eq:deltawdiss}
	\frac{\delta\omega}{\omega_{0,I}} &=  6  \lambda \Im{C^{(1)}}|\mathcal{A}|^2  - 12\lambda^2  \left(\Im{\hat{C}^{(2)}} + 3 \Re{C^{(1)}}\Im{C^{(1)}}\right) |\mathcal{A}|^4~.
	\ealigned{}
\end{align}
Eq. \eqref{eq:ampdiss} clearly shows that the saturation of the amplitude occurs when
\begin{align}\label{eq:77}
    |\mathcal{A}|^4 = -\frac{1}{12 \lambda^2 C^{(2)\diss}}~,
\end{align}
which takes the same form as Eq.~\eqref{eq:saturation}, but $\hat C^{(2)}$ is replaced with the dissipative part $C^{(2)\diss}$. As seen trivially from Eq.\eqref{eq:ampdiss}, this saturation occurs irrespective of the sign of $\lambda$.

However, we should be careful that the obtained saturation is real or not. The validity of our approximation would be examined by checking if the non-linearity is already important or not at the saturation amplitude. For this purpose, we can compare the two terms in the right hand side of Eq.~\eqref{eq:deltawdiss}. If the higher order term dominates, {\it i.e.}, if 
\begin{equation}\label{onemorecriterion}
	\left|\frac{\Im{\hat{C}^{(2)}} + 3 \Re{C^{(1)}}\Im{C^{(1)}}}
	 {\sqrt{3}\Im{C^{(1)}}|\sqrt{C^{(2)\diss}}}\right|\gtrsim 1\,,  
\end{equation}
it would be a sign of breakdown of the perturbative expansion. 

The above arguments tell us the qualitative behavior of the axion cloud and show the conditions for the saturation to occur in the minimal subtraction scheme and dissipative scheme. We will see in the next section that both schemes fail to show the existence of saturation. We understand that these arguments are still incomplete to conclude the absence of saturation, because there might be some other RG schemes that realize the saturation. However, we think that the absence of saturation in both distinctive schemes seems to be a strong evidence. One might be interested in up to which epoch our approximate treatment can be justified. To see it, we look at the size of the amplitude $\lambda |\mathcal{A}|^2$ and the relative magnitude of the frequency shift $\delta \omega/\omega_{0,R}$. As long as the perturbation theory is valid, these quantities must stay small. 

\section{Numerical results}\label{section:4}

In this section, we present the numerical calculation of the amplitude equation. In section \ref{sec:4.1}, we concentrate on the attractive self-interaction and in section \ref{sec:4.add}, we comment on the repulsive interaction. We used {\it Mathematica} for numerical calculation. 

\subsection{An example of time evolution of the amplitude equation}\label{sec:4.1}\mbox{}

Here, we present the time evolution of the axion cloud with the fastest growing mode. First, we will solve the RG equation numerically with the minimal subtraction scheme and the dissipative scheme, then we compare the two schemes to see whether the results change. After examples of the time evolution, we survey the $(\mu M, a/M)$ plane to examine whether there exist some parameter region in which the instability terminates.

\subsubsection{Examples of time evolution}\label{sec:4.1.1}

As an example, we present the time evolution of the frequency shift $\delta\omega$ in the case when the mass of the axion and the spin parameter of the central black hole are $\mu M = 0.42$ and $a/M = 0.99$, respectively. We consider the cloud evolved from a single mode with $l_0 = m_0 = 1$. With lowest overtone number $n_0 = 0$, this parameter set gives the growth rate of the axion cloud as 
\begin{align}
M\omega_I \sim 1.5038\times 10^{-7}~,
\end{align}
which is nearly the maximum and thus most interesting for realistic observations. 

\begin{figure}[h]
 \begin{tabular}{cc}
 \begin{minipage}[t]{0.5\hsize}
	\centering
	\includegraphics[keepaspectratio,scale=0.4]{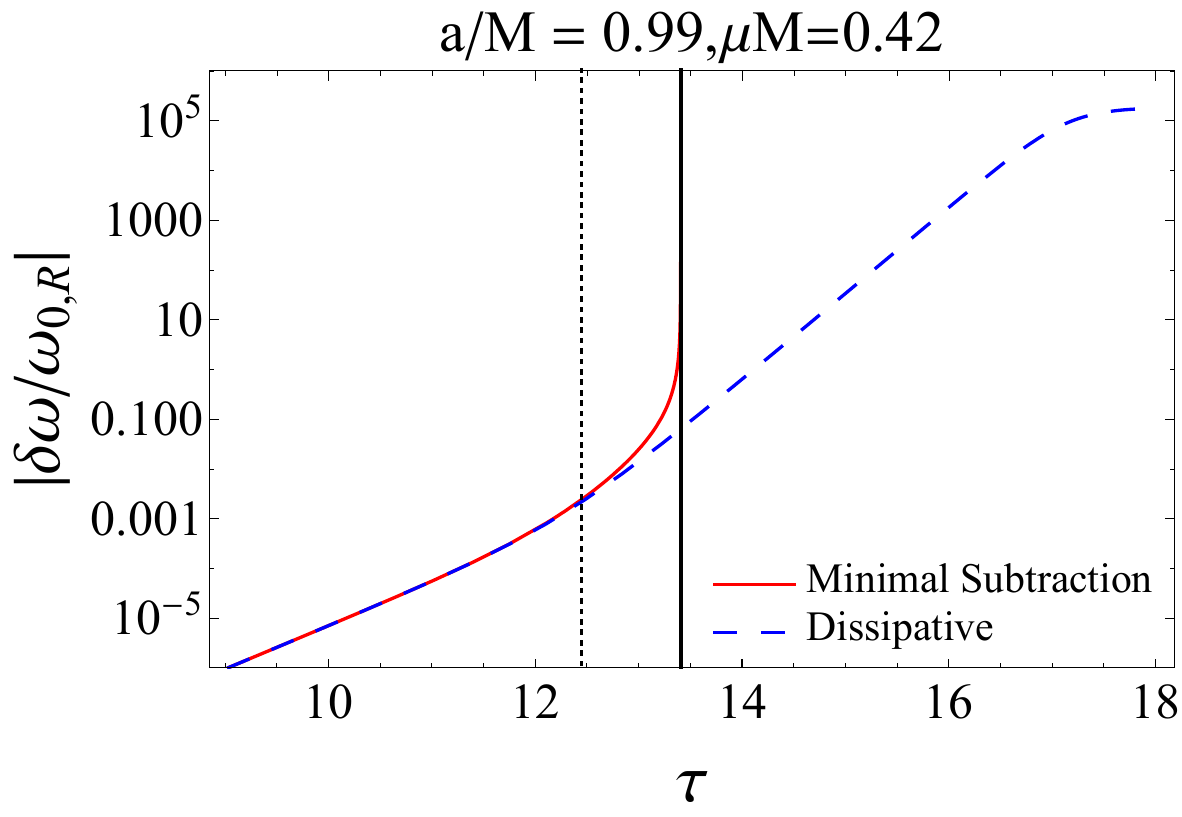}
\end{minipage} &
\begin{minipage}[t]{0.5\hsize}
        \centering
	\includegraphics[keepaspectratio,scale=0.55]{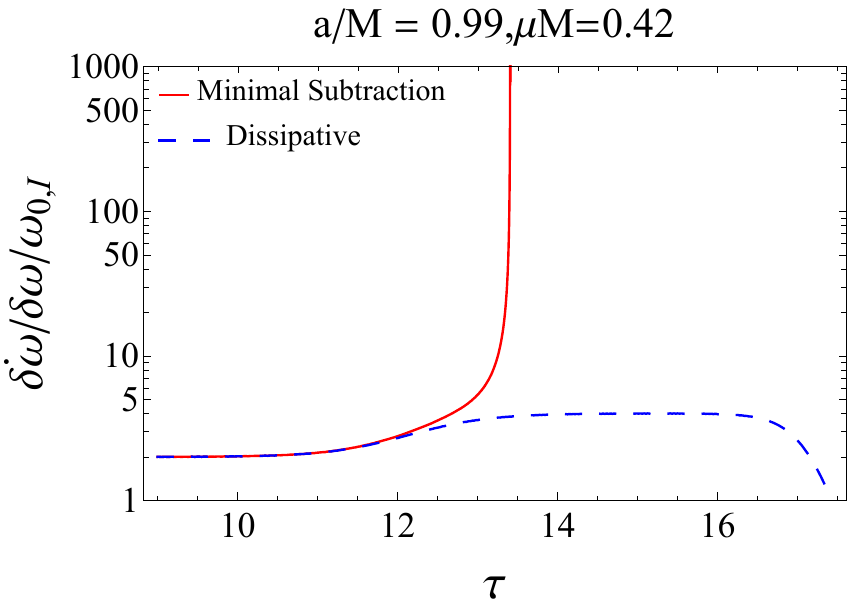}
\end{minipage}
\end{tabular}
	\caption{(Left) The red solid line and blue dashed line show time evolution of the amplitude for $l_0 = m_0 = 1,n_0=0, \mu M = 0.42, a/M = 0.99$ with minimal subtraction scheme and dissipative scheme, respectively. The horizontal axis is the normalized time $\tau:=\omega_{0,I} t$. The black dotted line shows the time when relative difference of the frequency shift in two schemes is 10\%, which we adopt for the criterion of the validity of our solution. The black solid line shows the time when the blow-up of the frequency shift occurs in the minimal subtraction scheme. (Right) The red and blue dashed lines show the inverse time scale of the axion cloud evolution $\dot \delta\omega/\delta\omega$ normalized by $\omega_{0,I}$ in the minimal subtraction scheme and the dissipative scheme, respectively.}
	\label{fig:3}
\end{figure}

\begin{table}[H]
  \caption{Numerical values of $C^{(1)}_{l_0m_0\omega_0}$ , $\hat{C}^{(2)}_{l_0m_0\omega_0}$ and $C^{(2)\diss}_{l_0m_0\omega_0}$ for the $l_0 = m_0 = 1,n_0=0, \mu M = 0.42, a/M = 0.99$ case.}
  \label{table:renomcoeef}
  \centering
  \begin{tabular}{cccc}
    \hline
    $C^{(1)}_{l_0m_0\omega_0}$ &$\hat{C}^{(2)}_{l_0m_0\omega_0}$ &$C^{(2)\diss}_{l_0m_0\omega_0}$\\
    \hline \hline
    $-7.82\times 10^{-3} - 2.33 \times 10^2 i$ & $1.54 \times 10^{-4} + 8.10 \times 10 i$ & $-1.73\times 10^{-10}$\\
    \hline
  \end{tabular}
\end{table}

In Table \ref{table:renomcoeef}, the calculated numerical coefficients that appear in the renormalization group equation is shown. We observe that the real parts of $C^{(1)}$ and $\hat{C}^{(2)}$ are both suppressed compared to its imaginary part, as is expected from the order counting with respect to $\omega_{0,I}$.

In Fig.~\ref{fig:3}, the time evolution tracks of the frequency shift $\delta\omega$ in the two schemes introduced in sec.~\ref{sec:3.1.3} are shown. As a representative case, we adopt the set of parameters given by $l_0 = m_0 = 1,n_0=0, \mu M = 0.42$ and $a/M = 0.99$. 
The horizontal axis is the normalized time 
\begin{equation}
     \tau:=\omega_{0,I} t.
\end{equation} 
The frequency shift blows up in the minimal subtraction scheme. This directly indicates the breakdown of the perturbation theory eventually occurs. On the other hand, saturation occurs in the dissipative scheme, as expected by construction. However, the fractional frequency shift $\delta\omega/\omega_{0,R}$ is already larger than unity when the saturation occurs. Therefore, although the qualitative behavior seems different, both schemes indicate the breakdown of the perturbation.

In the small amplitude regime, non-linearity is small and the frequency shift stays small. In this regime, the difference between the two schemes is negligibly small. However, eventually, non-linearity starts to affect the evolution. In the present case, non-linearity makes the cloud more unstable, and finally, the frequency shift gets large. This is the consequence of $\Re{\hat{C}_{l_0m_0\omega_0}^{(2)}} > 0$ in the minimal subtraction scheme and that of the condition~\eqref{onemorecriterion} in the dissipative scheme, as noted in the previous section.

Positivity of $\Re{\hat{C}_{l_0m_0\omega_0}^{(2)}}$ could be understood as follows. Note that the $m = m_0$ mode behaves as an ``almost'' bound state as we have mentioned below Eq.~\eqref{renomfirstorder} (we used ``almost'' because the boundary condition at the horizon is different from the bound state). Owing to the attractive nature of the axion potential (see Eq.~\eqref{potential}), the interaction energy of the axion cloud is negative and quartic in amplitude ${\cal A}$. This negative potential energy accelerates the increase of the amplitude, which results in the positive sign of $\Re{\hat{C}_{l_0m_0\omega_0}^{(2)}}$.  

Since our calculation is based on the perturbation theory, time evolution in the large frequency shift regime is not trustable. One might be interested in to what extent our approximate solution can be used as the initial condition for the dynamical simulation. For this purpose, we estimate the time scale of the successive non-linear evolution beyond the weakly non-linear regime in the minimal subtraction scheme. As shown in Fig.~\ref{fig:3} as a dotted vertical line, the time beyond which our renormalized solution cannot be trusted is around $\tau \sim 12$, where we adopt the criterion for the breakdown of perturbative solution that  the difference in the frequency shifts in two schemes is greater than 10 \%. Also, the time when the blow-up of the amplitude occurs is around $\tau \sim 13.5$ (the solid vertical line in Fig.~\ref{fig:3}). Therefore, the required time for the cloud to blow up after the perturbative approximation breaks down is estimated to be $\mathcal{O}(1)$ in the unit of the normalized time $\tau$. This indicates that tracking the succeeding non-linear evolution by a numerical simulation with the initial data provided by the current method would still be computationally expensive.

On the other hand, the time scale of the evolution $\dot{\delta \omega}/\delta \omega/\omega_{0,I}$ is still $\mathcal{O}(\omega_{0,I}^{-1})$ during the above intermediately non-linear regime (see the right panel of Fig.~\ref{fig:3}). This is still much larger than the dynamical time scale $\omega_{0,R}^{-1}$. Hence, the adiabatic approximation is still valid at this epoch.

\begin{figure}[t]
    \centering
     \begin{tabular}{cc}
 \begin{minipage}[t]{0.5\hsize}
	\centering
	\includegraphics[keepaspectratio,scale=0.35]{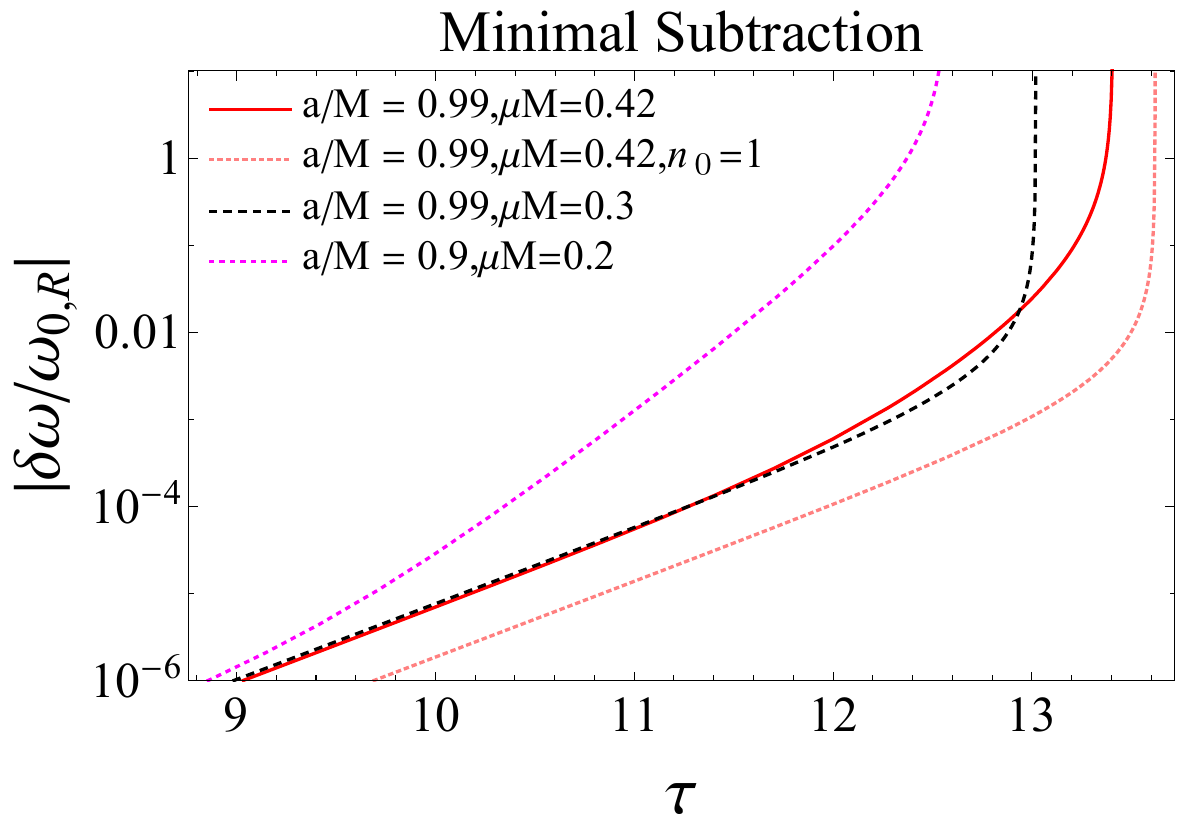}
\end{minipage} &
\begin{minipage}[t]{0.5\hsize}
        \centering
	\includegraphics[keepaspectratio,scale=0.35]{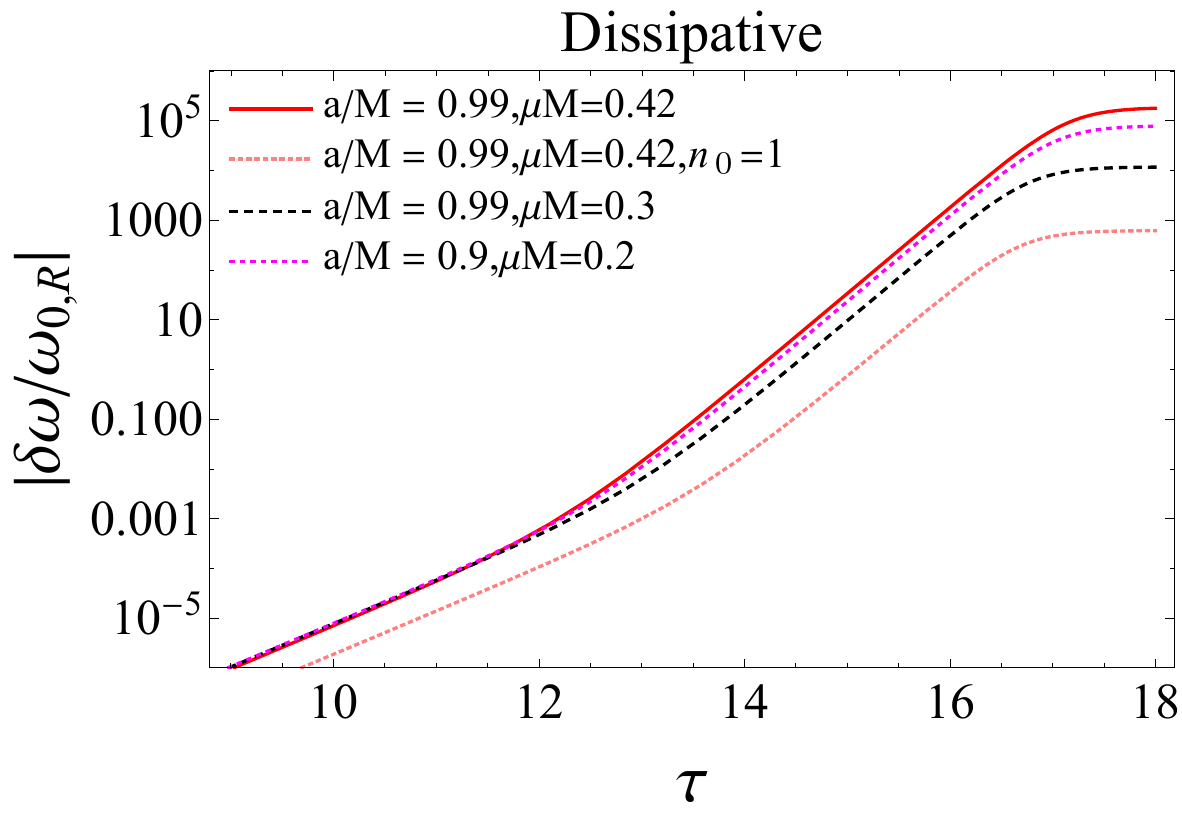}
\end{minipage}
\end{tabular}
    \centering
     \begin{tabular}{cc}
 \begin{minipage}[t]{0.5\hsize}
	\centering
	\includegraphics[keepaspectratio,scale=0.35]{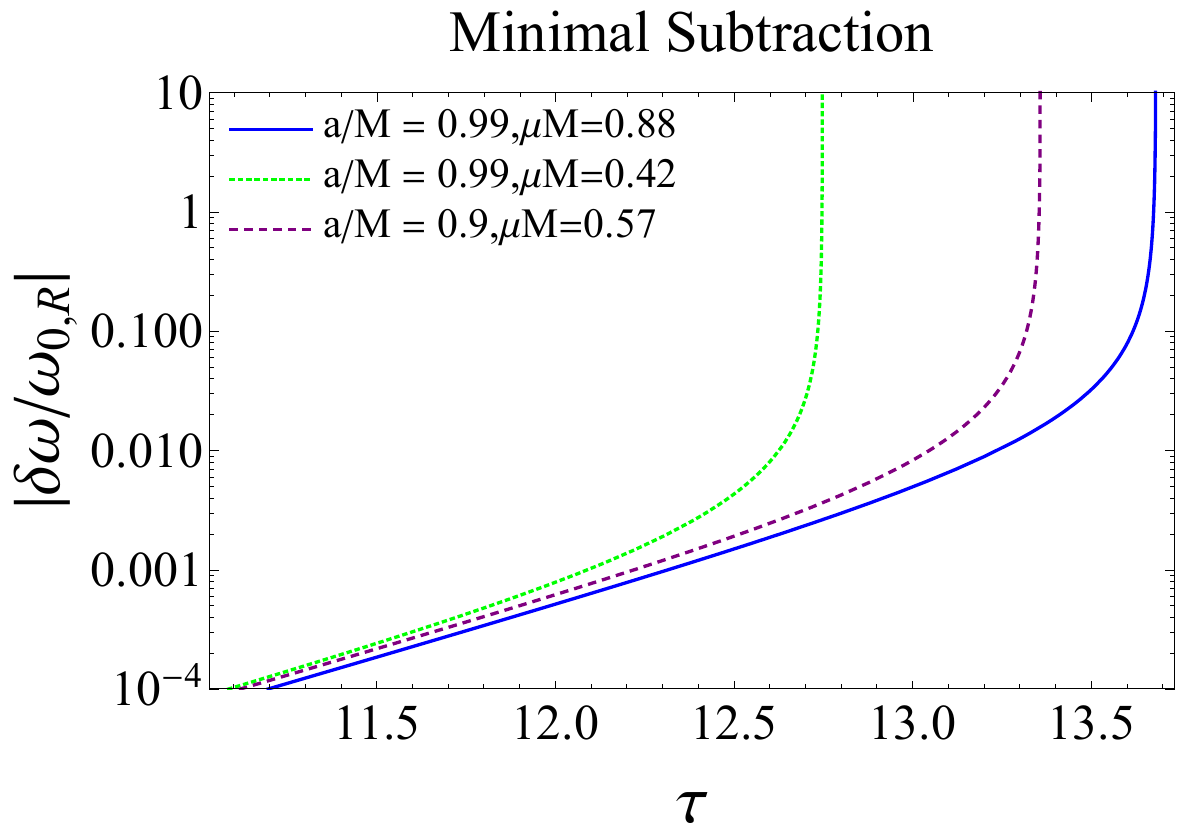}
\end{minipage} &
\begin{minipage}[t]{0.5\hsize}
    \centering
	\includegraphics[keepaspectratio,scale=0.35]{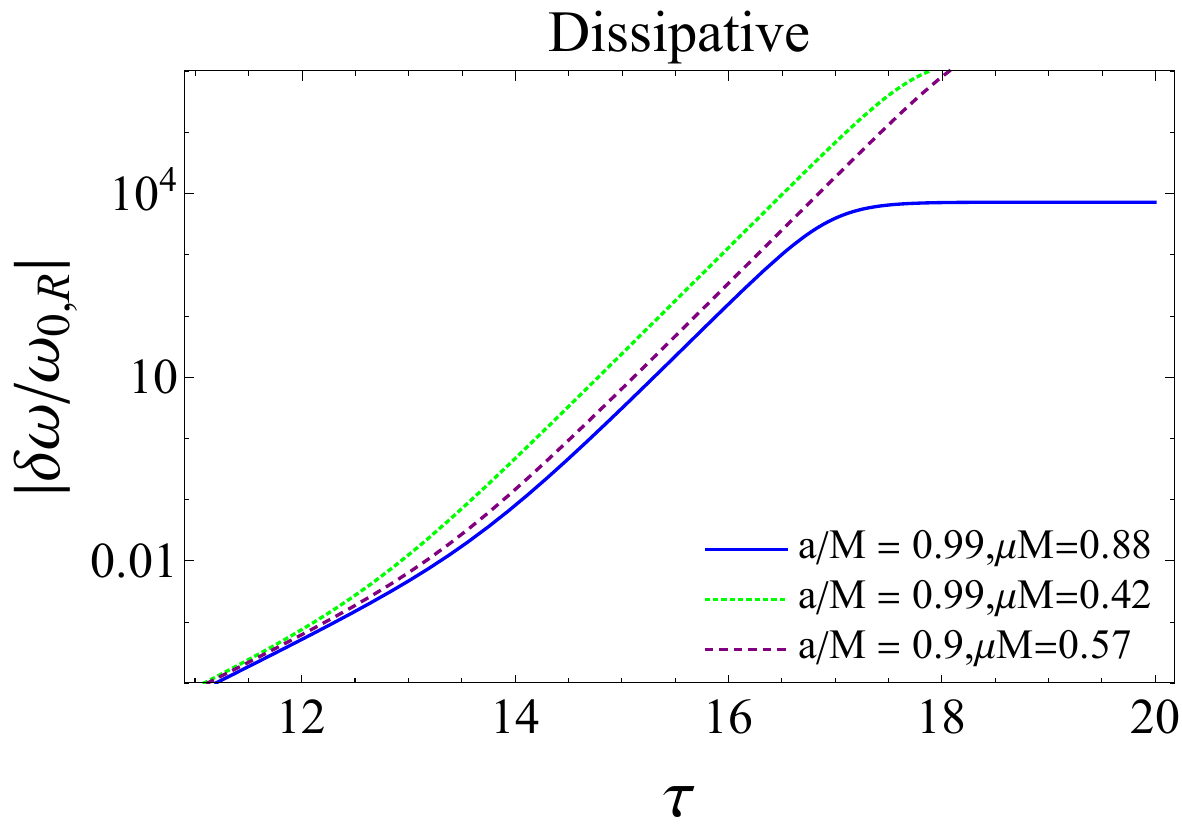}
\end{minipage}
\end{tabular}
	\caption{Left and right panels show the evolution of the frequency shift with the minimal subtraction scheme and dissipative scheme, respectively. Top two panels show $l_0=m_0=1$ modes and Bottom two panels show $l_0=m_0=2$ modes. Each line corresponds to the different $(a/M,\mu M,l_0,m_0,n_0)$ modes. The red solid line, red dotted line, black dashed line, pink dotted, blue solid line, green dotted line, and black purple line corresponds to $(0.99,0.42,1,1,0)$, $(0.99,0.42,1,1,1)$, $(0.99,0.3,1,1,0)$, $(0.9,0.2,1,1,0)$, $(0.99,0.88,2,2,0)$, $(0.99,0.42,2,2,0)$, and $(0.9,0.57,2,2,0)$ modes, respectively.}
	\label{fig:4}
\end{figure}

We changed the axion mass $\mu M $ and the spin parameter $a/M$ to see if the breakdown of the renormalized perturbation before the saturation of amplification is an universal feature of the relativistic cloud. Figures~\ref{fig:4} show the evolution of the frequency shift for $l_0 = m_0 = 1,2$ modes adopting the minimal subtraction scheme and dissipative scheme. Table.~\ref{table:calculatedmodes} shows the frequency and the growth rate of these modes as well as $\Re{C^{(1)}}$, $\Re{\hat{C}^{(2)}}$ and $C^{(2)\diss}$. We confirm that the perturbative expansion breaks down as the frequency shift gets large also in these cases. 

\begin{table}[H]
  \caption{Growth rate of the mode we have calculated. We also present $\Re{C^{(1)}_{l_0m_0\omega_0}}$, $\Re{\hat{C}^{(2)}_{l_0m_0\omega_0}}$ and $C^{(2)\diss}_{l_0m_0\omega_0}$.}
  \label{table:calculatedmodes}
  \centering
  \begin{tabular}{lccccc}
    \hline
    $(a/M,\mu M, l_0, m_0,n_0)$  & $M \omega_R$  &  $M \omega_I$ & $\Re{C^{(1)}_{l_0m_0\omega_0}}$& $\Re{\hat{C}^{(2)}_{l_0m_0\omega_0}}$ & $C^{(2)\diss}_{l_0m_0\omega_0}$\\
    \hline \hline
    (0.99,0.42,1,1,0)  & 0.4088  & $1.5038\times 10^{-7}$ &$-7.82\times 10^{-3}$ & $1.54 \times 10^{-4}$ & $-1.73\times 10^{-10}$\\
    (0.99,0.42,1,1,1)  & 0.4151 & $5.3621\times 10^{-8}$ & $-1.99\times 10^{-3}$ & $5.38\times 10^{-6}$ & $-1.23\times 10^{-11}$\\
    (0.99,0.3,1,1,0)  & 0.2963  & $2.6806\times 10^{-8}$ & $-2.95\times 10^{-3}$ & $1.99\times 10^{-5}$ & $-2.20\times 10^{-11}$ \\
    (0.9,0.2,1,1,0)  & 0.1990  & $1.5562\times 10^{-9}$ & $-1.17\times 10^{-3}$ & $3.34 \times 10^{-6}$ & $-2.18\times 10^{-13}$\\
    (0.99,0.88,2,2,0)  & 0.8508  & $2.7628\times 10^{-8}$ & $- 1.11\times 10^{-2} $&$ 2.63\times 10^{-4}$ & $-2.12 \times 10^{-9}$\\
    (0.99,0.42,2,2,0)  & 0.4156  & $6.3702\times 10^{-11}$ & $-1.53\times 10^{-3}$ & $4.99\times 10^{-6}$ & $-3.38 \times 10^{-14}$\\
    (0.9,0.57,2,2,0)  & 0.5585  & $1.0303\times 10^{-9}$ & $-3.14\times 10^{-3}$ & $1.02\times 10^{-7}$ & $-6.75 \times 10^{-14}$ \\
    \hline
  \end{tabular}
\end{table}

\subsubsection{Survey of the parameter space}\label{sec:4.2}

From the results of the previous section, we find that the growth of relativistic axion clouds does not seem to saturate at least without entering the strongly non-linear regime, which is beyond the scope of our current approximation. In this section, we investigate whether this is really a generic feature of the self-interacting axion cloud around a rotating black hole or not. We present our results, focusing on the $(l_0,m_0) = (1,1)$ modes here.

We see that the termination of the growth by saturation in the weakly non-linear regime is possible only when (1) $\Re{\hat{C}^{(2)}}$ is negative in the minimal subtraction scheme, or (2) the size of the left hand side of Eq.\eqref{onemorecriterion} is smaller than unity in the dissipative scheme. The second condition can be achieved when $|C^{(2)\diss}|$ is sufficiently large. In other words, in both cases the dissipative effect must overwhelms the acceleration of the growth of cloud due to the attractive self-interaction. Thus, we explore the behavior of $\Re{\hat{C}^{(2)}}$, the left hand side of Eq.~\eqref{onemorecriterion}, and $|C^{(2)\diss}|$ on the $(a/M, \mu M)$ plane. 

\begin{figure}
    \centering
	\includegraphics[keepaspectratio,scale=0.5]{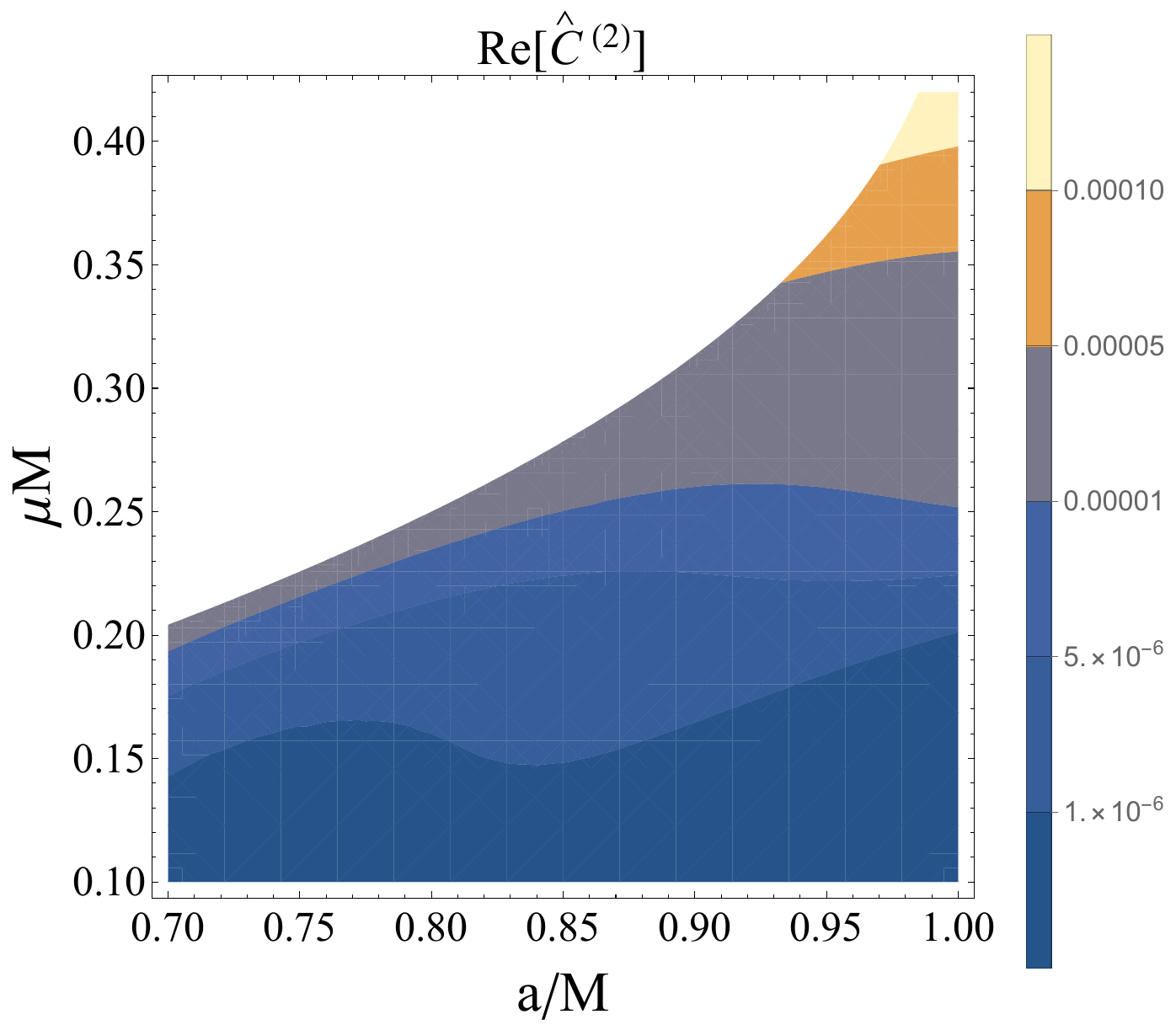}
	\includegraphics[keepaspectratio,scale=0.5]{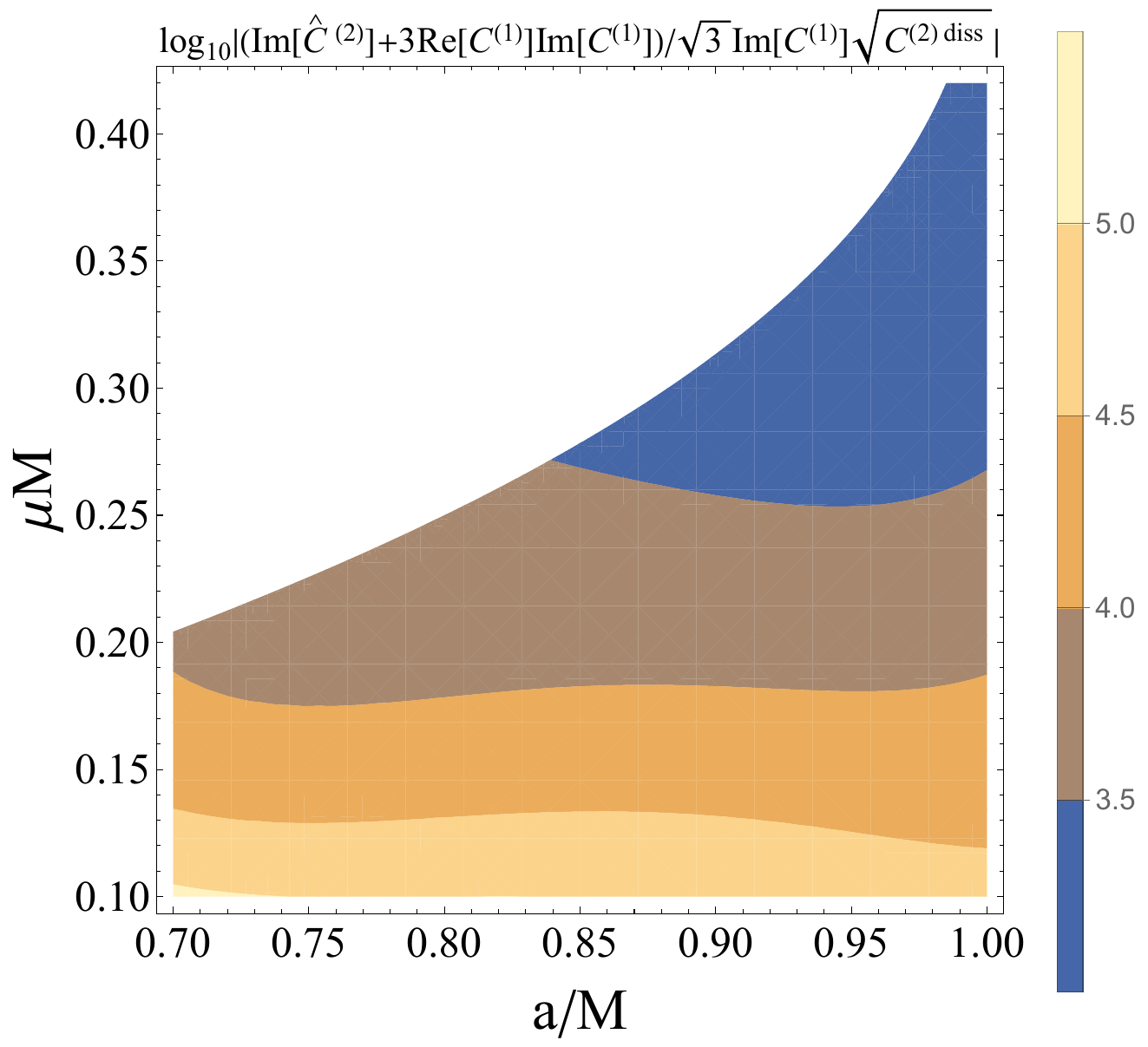}
	\caption{Top and bottom panels show the behavior of the $\Re{\hat{C}^{(2)}}$ and $\log_{10}(\rm{L.H.S.\ of\ Eq.\eqref{onemorecriterion}})$ on the $(a/M,\mu M)$ plane, respectively. The boundary of the region is given by the superradiance condition $\mu M < m \Omega_H$.}
	\label{fig:7}
\end{figure}

\begin{figure}
    \centering
    \includegraphics[keepaspectratio,scale=0.5]{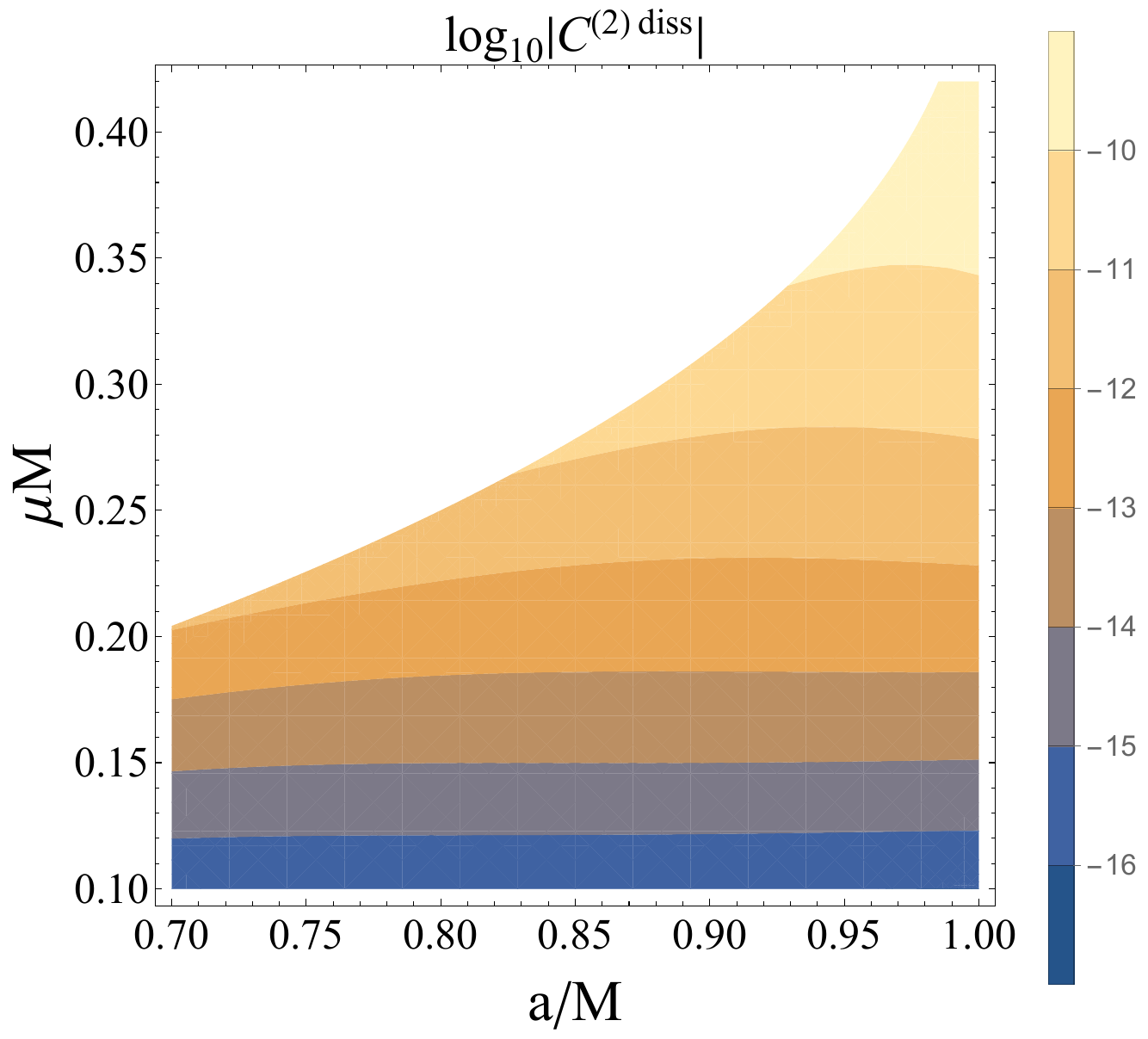}
 	\includegraphics[keepaspectratio,scale=0.5]{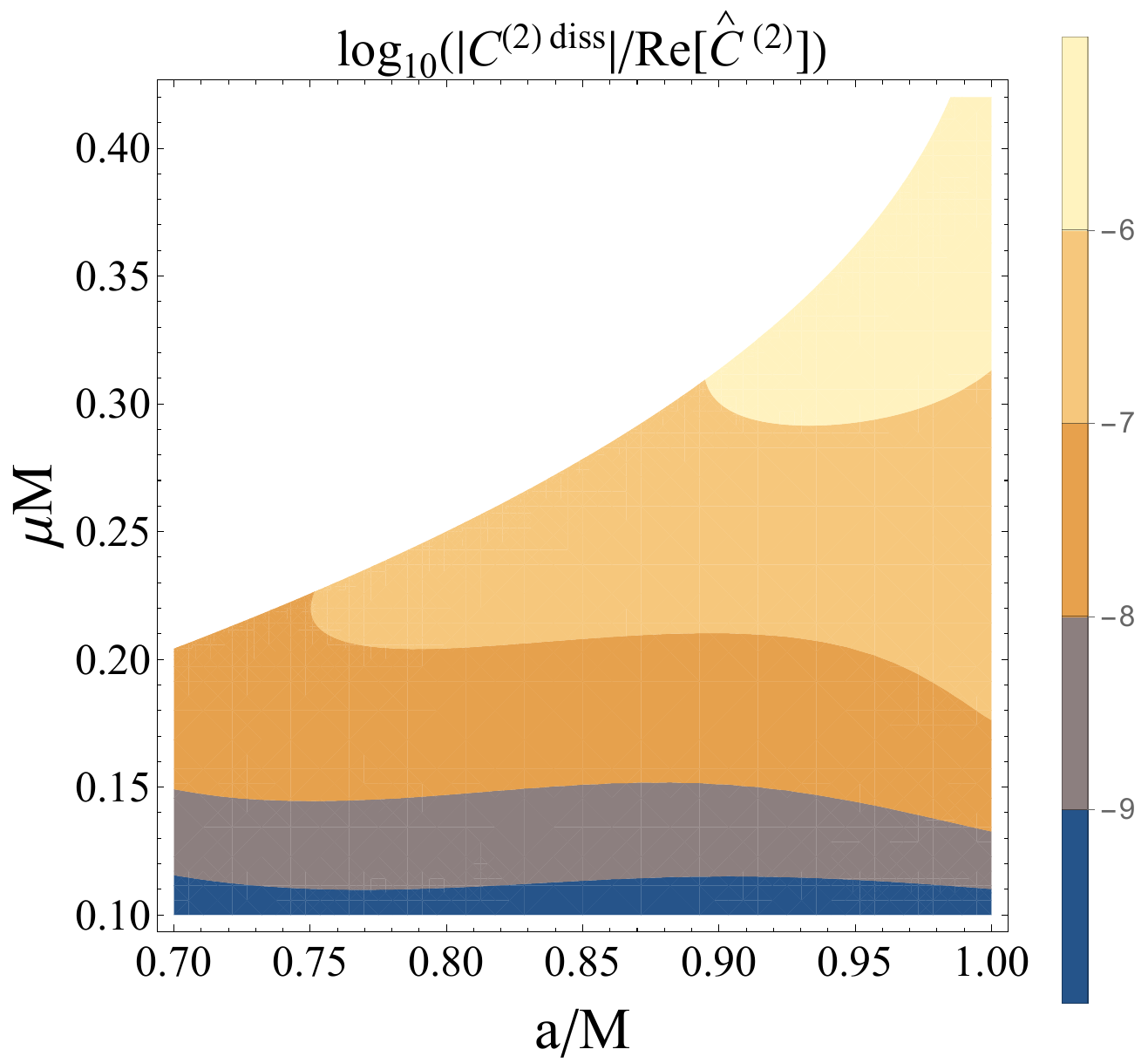}
    \caption{Top and bottom panel shows the behavior of the left hand side of $\log_{10}|C^{(2)\diss}|$ and $\log_{10}(|C^{(2)\diss}|/\Re{\hat{C}^{(2)}})$ on the $(a/M,\mu M)$ plane, respectively. The boundary of the region is given by the superradiance condition $\mu M < m \Omega_H$.}
    \label{fig:add2}
\end{figure}

In Fig.~\ref{fig:7}, we show $\Re{\hat{C}^{(2)}}$ and the left hand side of Eq.~\eqref{onemorecriterion} on the $(a/M, \mu M)$ plane. We see that $\Re{\hat{C}^{(2)}}$ is always positive and 
the left hand side of Eq.\eqref{onemorecriterion} is much larger than unity. Also, we show the behavior of $|C^{(2)\diss}|$ and $|C^{(2)\diss}|/\Re{\hat{C}^{(2)}}$ on the $(a/M,\mu M)$ plane. We observe that $|C^{(2)\diss}|$ gets smaller for the smaller $a/M$ or $\mu M$, and hence $C^{(2)\diss} \ll \Re{\hat{C}^{(2)}}$ holds for any parameter set. We conclude that dissipation effect is relatively small. As a result, the inequality Eq.\eqref{onemorecriterion} holds.

In the non-relativistic region ($\mu M\ll 1$), although the cloud extends to a large radius, the dissipation due to the radiation to infinity gets less efficient. One may think that the multipole moments that become the source of radiation become large, 
and hence the radiative dissipation should be efficient. However, the wave length of the outgoing wave $\lambda_{out} = 1/(3\omega_0) \sim 1/(3\mu)$ is much smaller than the size of the cloud $r_c \sim M/(\mu M)^2$, especially in the non-relativistic regime. Therefore, because of the phase cancellation, only the part of the cloud close to the horizon contributes to the flux integral $j_{l3m_03\omega_0}$, which makes the dissipation to infinity inefficient. For this reason, $C^{(2)\diss}$ gets small in the non-relativistic regime. These results suggest that the growth of the cloud to a strongly non-linear regime is a generic feature of  self-interacting axion clouds. 

\subsection{Repulsive self-interaction}\label{sec:4.add}
Finally, we will briefly discuss the case of repulsive self-interaction. 
For negative $\lambda$, the time evolution tracks of $\delta\omega$ in the two schemes are shown in Fig.~\ref{fig:add}. Also in this case, the late time behaviors in the two schemes are different, which suggests the breakdown of the perturbative approach. In principle, there is a possibility that only the dissipative scheme maintains the validity of perturbative expansion. Even if we assume so, the significance of the non-linearity at the saturation amplitude can be investigated by examining the magnitude of $\delta\omega/\omega_{0,R}$ or the left hand side of \eqref{onemorecriterion}. 
As seen in the previous subsection, the inequality~\eqref{onemorecriterion} is always satisfied (see  Fig.~\ref{fig:add2}). This signifies that the breakdown of the perturbative expansion before the saturation occurs even in the repulsive self-interaction.

\begin{figure}
    \centering
    \includegraphics[keepaspectratio,scale=0.5]{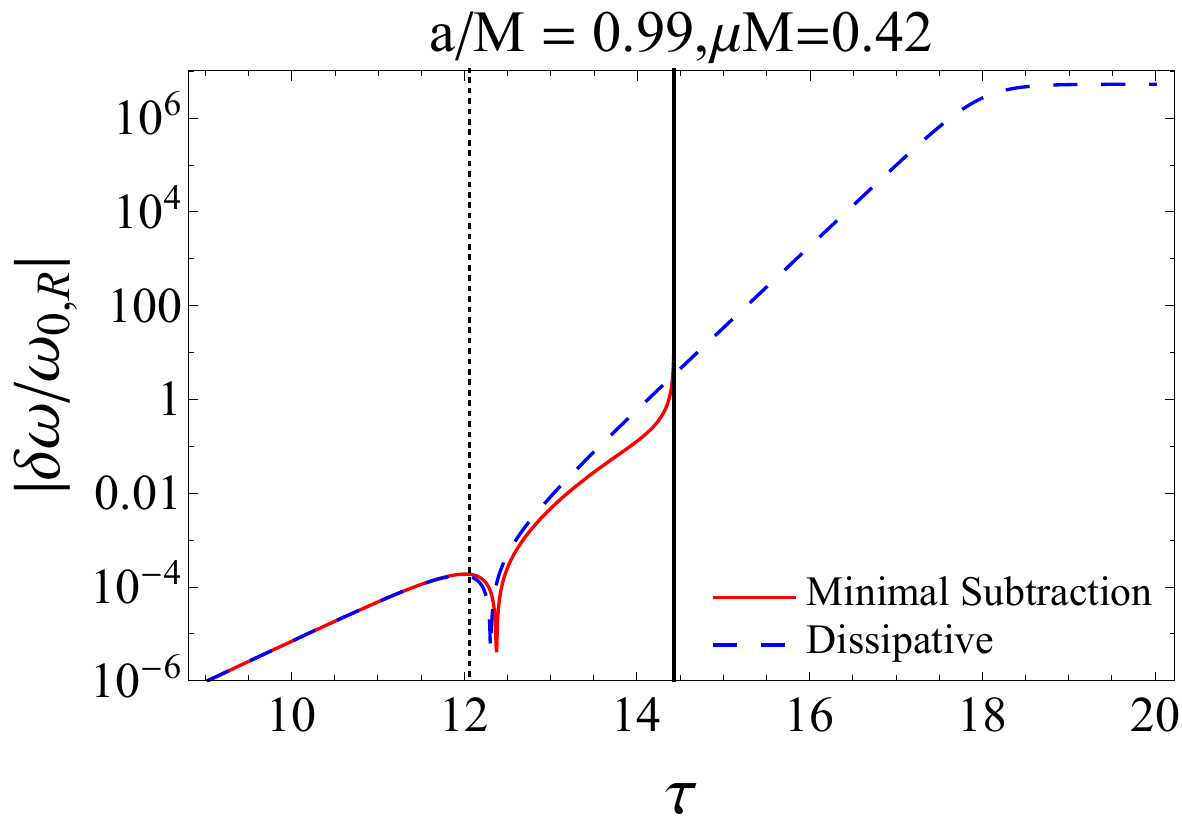}
    \caption{The red solid line and blue dashed line shows time evolution of the amplitude for $l_0 = m_0 = 1,n_0=0, \mu M = 0.42, a/M = 0.99,\lambda < 0$ with minimal subtraction scheme and dissipative scheme, respectively. The horizontal axis is the normalized time $\tau:=\omega_{0,I} t$.}
    \label{fig:add}
\end{figure}

\section{Comment on the multiple superradiant mode}\label{section:add}

Recently, \cite{Baryakhtar:2020gao} claimed that the interaction between multiple superradiant modes induced by self-interaction will terminate the growth of the cloud. Their calculation is based on the non-relativistic approximation ($\mu M_{BH} \ll 1$) and cannot tell whether this termination occurs in the relativistic regime ($\mu M_{BH} \gtrsim 1$). Since our formulation is applicable even in the relativistic regime, we can investigate this saturation also in this regime, although the discussion in this paper is restricted to the case with only a single superradiant mode. In this section, we briefly review the interaction between the multiple modes and then we combine the results in \cite{Baryakhtar:2020gao} and ours to estimate whether the saturation due to co-existence of multiple superradiant modes may  affect our analysis in the relativistic regime.

\begin{figure}
    \centering
    \includegraphics[keepaspectratio,scale=0.3]{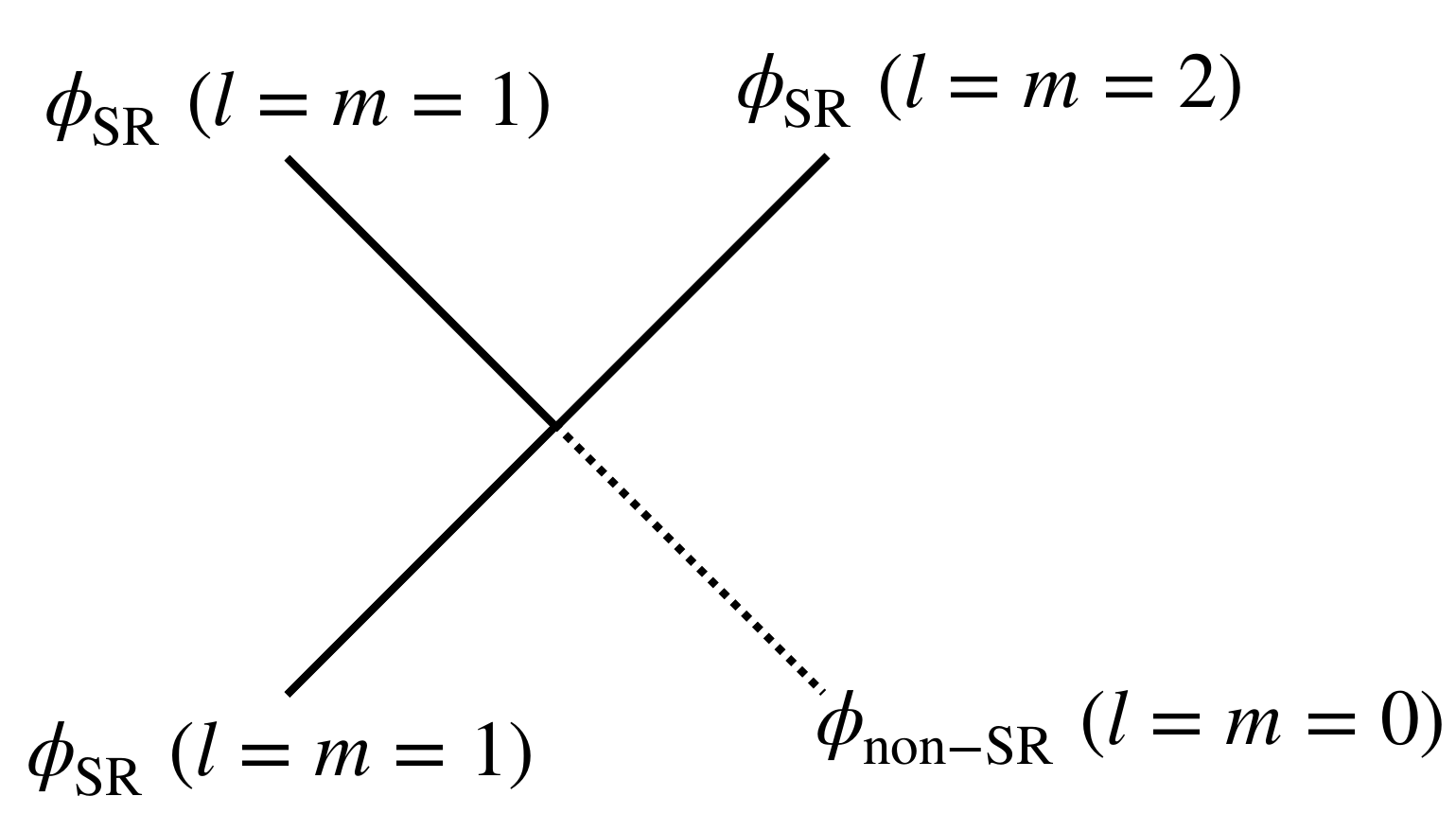}
    \caption{The diagram of the interaction between the bound state. The quantity $\phi_{\rm SR}$ and $\phi_{\rm non-SR}$ denotes the superradiant mode and non-superradiant mode of the axion, respectively. In this figure, we showed the case for the lowest angular momentum modes. This interaction is possible since $2\omega_{R}(l=m=1) - \omega_{R}(l=m=2) >0$ as shown in Table \ref{table:calculatedmodes}.}
    \label{fig:extra}
\end{figure}

If we consider the axion cloud with multiple superradiant modes, self-interaction will induce stimulated emission of non-superradiant mode \cite{Baryakhtar:2020gao} (see Fig. \ref{fig:extra}). Since the non-superradiant mode falls into the black hole, this process shrinks the axion cloud. Therefore, the stimulated emission will eventually balance with the superradiant growth, and the cloud becomes a quasi-equilibrium state. In \cite{Baryakhtar:2020gao}, the detailed analysis of this effect has been developed in the non-relativistic regime and they concluded that this quasi-equilibrium state will be realized in the weakly non-linear regime. 

However, in our opinion, it would be too early to conclude that the growth of clouds always terminate. First, their analysis is restricted to the non-relativistic regime and treats the interaction perturbatively. In addition, their calculations do not seem to consider the effect of acceleration of cloud instability due to the attractive self-interactions revealed in our study. Thus, we need a further investigation to understand the impact of the multiple modes on the evolution of the cloud.

\begin{figure}
    \centering
    \includegraphics[keepaspectratio,scale=0.5]{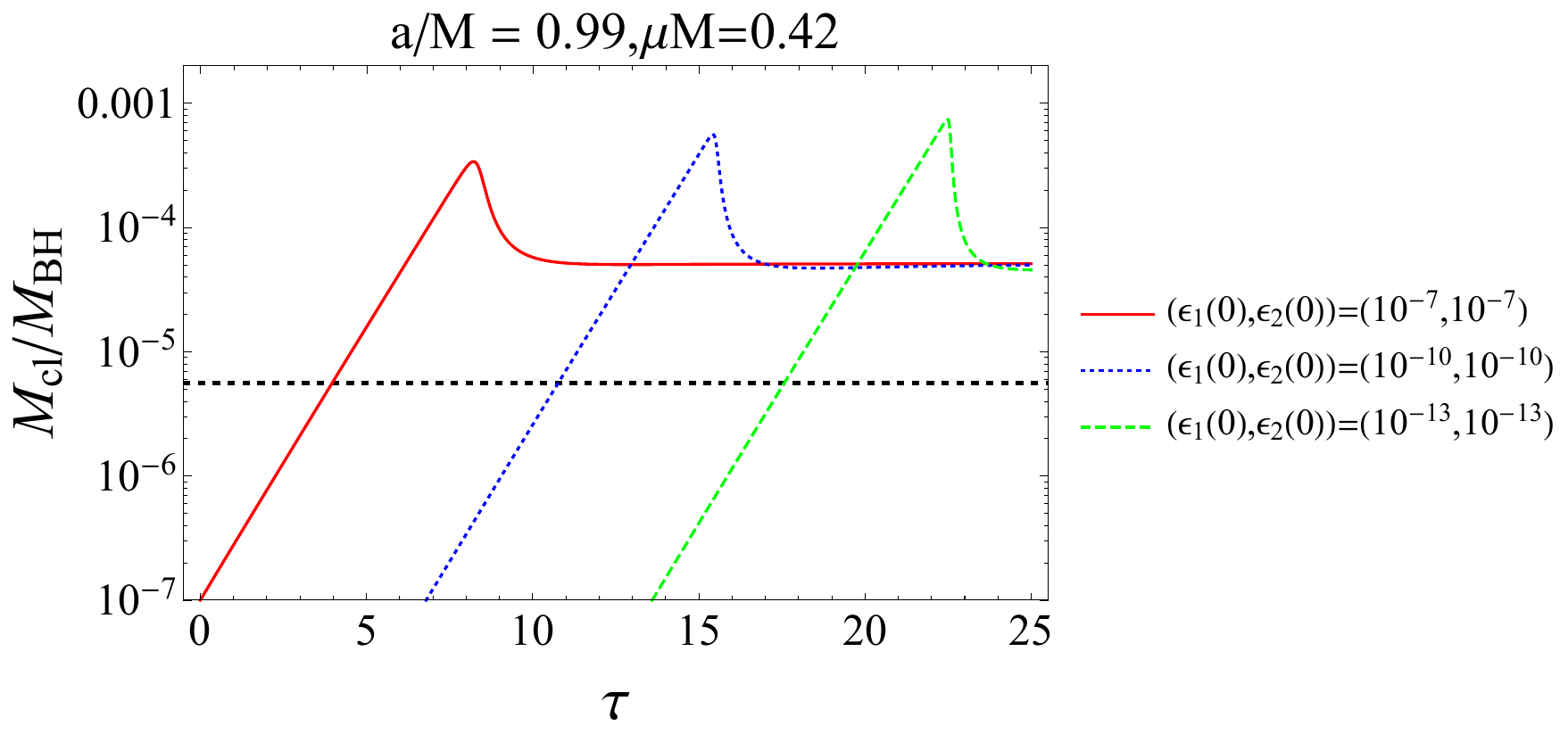}
    \caption{The red solid line and the blue dashed lines correspond to the evolution of the cloud mass, starting with a few different initial amplitudes, $(\epsilon_1(0),\epsilon_2(0)) = (10^{-7},10^{-7}),(10^{-10},10^{-10}),(10^{-13},10^{-13})$. Here, $\epsilon_i$ $(i = 1,2)$ are the cloud mass normalized by the black hole mass. The subscript $i$ discriminates the two superradiant modes. In this case, $i=1$ and $i=2$ correspond to the $l=m=1$ and $l=m=2$ modes, respectively. The horizontal black dashed line corresponds to the threshold value of the cloud mass for the reliable perturbative expansion (see Eq. \eqref{eq:add1}).}
    \label{fig:add}
\end{figure}

As a first step of this investigation, we estimate whether the saturation of a relativistic cloud with multiple modes occurs within the weakly nonlinear regime or not. Recall that our calculation can tell the breakdown of the perturbative treatment by examining the relative difference of two schemes (see Sec. \ref{sec:4.1.1}). For the concreteness, we fix the parameter set of the cloud to $(a/M_{BH},\mu M_{BH},l,m) = (0.99,0.42,1,1)$. Then, our calculation tells that when
\begin{align}\label{eq:add1}
    \frac{M_{cl}}{M_{BH}} \sim 5.6 \times 10^{-6} \left(\frac{F_a}{10^{16} {\rm GeV}}\right)^2~,
\end{align}
the breakdown of perturbation theory occurs. We approximated the mass of the axion cloud by
\begin{align}
    M_{cl} &\sim \mu |A|^2 r_{cl}^3~, & r_{cl} \sim \frac{l}{G\mu^2 M_{BH}}~.
\end{align}
Here, $r_{cl}$ is the size of the axion cloud. 

\begin{figure}
    \centering
    \includegraphics[keepaspectratio,scale=0.5]{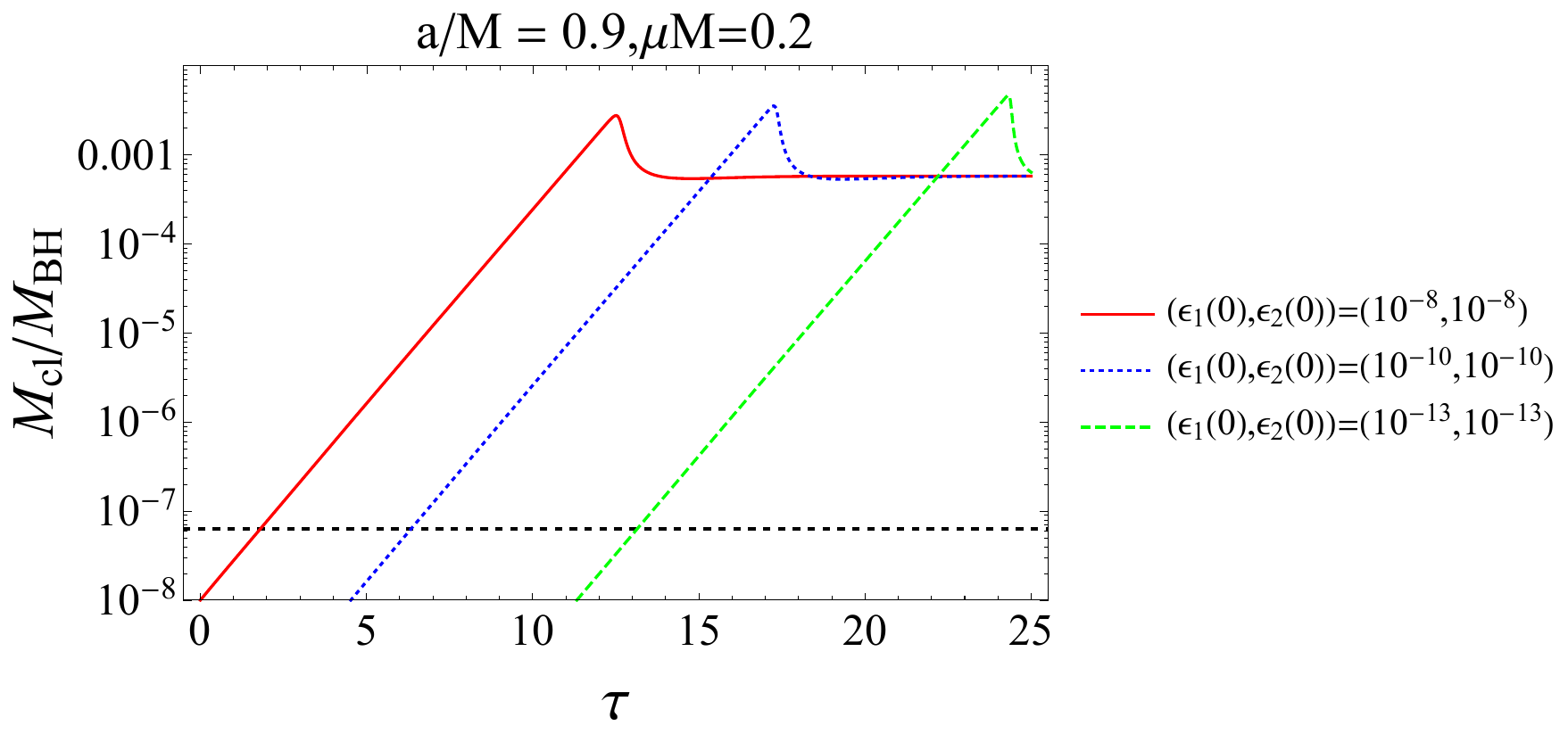}
    \caption{The same plot as Fig.~\ref{fig:add}, but  
    with $(a/M_{BH},\mu M_{BH},l,m) = (0.9,0.2,1,1)$. 
    }
    \label{fig:add3}
\end{figure}

By solving the evolution equations given in \cite{Baryakhtar:2020gao} (see Eqs. (34) and (35) in their paper, neglecting the effect of the scalar and the gravitational wave emission), we can tell how the mass of the axion cloud evolves under the presence of the second superradiant mode. In Fig.~\ref{fig:add}, we show an example of the evolution with a few different initial conditions. The initial stage of evolution follows an exponential growth obtained by linear analysis. When the cloud grows to some extent, the interaction shown in Fig. \ref{fig:extra} becomes dominant, and the growth of the cloud terminates. However, we observe that the perturbative treatment of the self-interaction breaks down before the saturation. Consequently, a more sophisticated approach to deal with clouds in highly nonlinear regions is needed to determine whether their growth is saturated by the effect of coexistence of multiple modes. We also confirmed that this is true for the mildly relativistic case, $(a/M_{BH},\mu M_{BH},l,m) = (0.9,0.2,1,1)$ (see Fig. \ref{fig:add3}).

\section{Summary and Discussion}
\label{section:5}

Axion clouds around spinning black holes will induce interesting phenomena and allow us to test the existence of new particles via gravitational waves or other astrophysical probes. For further investigation of this possibility, understanding the precise dynamics of the axion cloud is important. In this paper, We examined the impact of the axion self-interaction on the dynamics of an axion cloud composed of a single superradiant mode by applying the dynamical renormalization group method. In particular, we focused on the possibility that the energy loss through the scalar field emission balances with the energy input due to the superradiance to realize a saturated quasi-stationary configuration. If the saturation occurs, explosive phenomena such as bosenova, which might be relevant for the direct observation, would not occur.  

There are two types of effects caused by the self-interaction. One is the radiation of the axion field which dissipates the energy of the axion cloud. This effect is often ignored in the literature and it was one of our aims to clarify whether this effect saturates the growth of an axion cloud or not. Our calculation indicates that this dissipation does not work efficiently in any case, especially since the dissipation effect gets less important in the non-relativistic regime, $\mu M \to 0$ and $a/M \to 0$. The other type of effect is that the attractive self-interaction lowers the energy of the cloud to accelerate the instability. 

Our results suggest that the saturation of the growth does not occur in the weakly non-linear regime, as long as we consider the $\phi^4$ type self-interaction. More precisely, the perturbative expansion breaks down before the dissipation starts to affect the cloud evolution. Inefficient dissipation due to the axion radiation to infinity is the main reason for this conclusion.

However, our analysis does not indicate that the explosive phenomena like a bosenova and hence bursts of gravitational waves always occur. Several aspects, which have not been taken into account in our current analysis, may terminate the instability. First of all, we cannot say anything about the evolution after the perturbative approach breaks down. 
In this paper, we truncated the cosine type potential at the $\phi^4$ term. When the cloud gets denser, higher order terms in the potential come into play. The non-linear effects induced by the higher order terms in the cosine type potential may change the evolution. 

Another possibility is the contamination of non-superradiant modes (such as $m=-1$ modes), which can dissipate energy efficiently by falling into the black hole. 
Within our current setup, there is no room for such modes to appear because the frequencies of the possibly excited modes are restricted to $m\omega_0$, which all satisfy the condition for the superradiance. This situation is likely when we consider an adiabatic growth of the cloud from the fastest growing mode. However, once an explosive phenomenon occurs, the state with plural superradiant modes excited will be realized. In such a situation, the self-interaction can induce non-superradiant excitations. 

Then, the cloud may develop into a quasi-stationary state, in which superradiant modes extract the mass and angular momentum of the black hole, while non-superradiant modes give them back to the black hole.  If such a quasi-stationary state is realized with a relatively small amplitude, {\it i.e.}, much smaller than the case without self-interaction, in which the gravitational wave radiation balances with the superradiant energy extraction, the spin parameter of the central black hole will change much more slowly. Then, the constraint from the population of black holes on the Regge plane~\cite{Ng:2020ruv,Stott:2020gjj,Zu:2020whs},
will be modified from the one without the self-interaction~\cite{Mathur:2020aqv}.

The impact of simultaneous excitation of plural superradiant modes on the evolution of an axion cloud has already been pointed out in Ref.~\cite{Ficarra:2018rfu, Baryakhtar:2020gao} and they claim that the instability is saturated by the interaction between the modes. We have estimated whether this saturation occurs in weakly non-linear regime or not by using our evolution equation, and found that it is not the case. Of course, our formulation presented in this paper applies only to the case with a single superradiant mode, and thus our estimation might be altered by considering the interaction between multiple modes. To do so, we can simply extend our formulation to the case with multiple superradiant modes. Then, one can calculate the evolution of a cloud correctly capturing the existence of the multiple superradiant modes in a wide parameter region. We leave this extension to the future work.

To tell whether bosenovae happen or not in the models with an axion field having an appropriate mass, we need an alternative numerical analysis method that can track the adiabatic evolution of the cloud in the strongly non-linear regime without the truncation of the potential. Full numerical simulation would be one option to treat such a problem~\cite{Yoshino:2012kn,Yoshino:2015nsa,East:2017ovw,Santos:2020pmh,Herdeiro:2014goa}, although the computational cost is high. Our method will give a better initial data than just assuming a configuration in an ad-hoc manner, when we consider to start the simulation with a mildly large amplitude to reduce the computational cost. 

\section*{Acknowledgements}
This work was supported by JSPS KAKENHI Grant Number JP17H06358 (and also JP17H06357), \textit{A01: Testing gravity theories using gravitational waves}, as a part of the innovative research area, ``Gravitational wave physics and astronomy: Genesis'', and also by JP20K03928. 

\appendix

\section{Renormalization Group method for differential equation}
\label{App:A}

The RG method is a resummation technique to eliminate a secular term in perturbative solution~\cite{Chen:1994zza,10.1143/PTP.94.503,Ei:1999pk}. Here, we give two examples that may help understanding the application of RG method to the axion cloud.

\subsection{Rayleigh-Schr\"odiner perturbation}

Here, we solve perturbation problem of the Schr\"odinger equation with RG method. Consider following equation
\begin{align}\label{1}
	\left(i\PD{t} - H_0\right)\psi(t,x) = \epsilon V(x) \psi(t,x)~,
\end{align}
with $\epsilon \ll 1$. 
We assume that $H_0$ is solved explicitly and has eigenvalue $E_n$ with eigenfunction $u_n$. Let us solve this equation perturbativly in $\epsilon$. If we write $\psi = \psi_0 + \epsilon \psi_1 + \dots$, then we obtain
\begin{align}
	\label{2}
	\left(i\PD{t} - H_0\right)\psi_0(t,x) &= 0~,\\
	\label{3}
	\left(i\PD{t} - H_0\right)\psi_1(t,x) &= V\psi_0~.
\end{align}
We take zeroth order solution as
\begin{align}
	\psi_{0,m}(t,x) &=  A(t_0)e^{-iE_m t}u_m(x)~.
\end{align}
Here, $A(t_0)$ is the amplitude of the wave function at $t=t_0$. We can solve the first order equation~\eqref{3} by using the Green's function of Eq.~\eqref{2}, which can be written by using mode function $u_n$ as
\begin{align}
	G(t,x;t',x') = \sum_n\int\frac{d\omega}{2\pi}\frac{u_n(x)u^*_n(x')}{\omega - E_n}e^{-i\omega(t-t')}~.
\end{align}
Then, the solution of Eq.~\eqref{3} is
\begin{align}
\baligned{}
	\label{eq4}
	\psi_1 &= A(t_0)\sum_{n\neq m}u_n(x)\left(\int dx'u^*_n(x')Vu_m(x')\right) e^{-iE_n t}\frac{e^{i(E_n-E_m)t}-e^{i(E_n-E_m)t_0}}{E_n - E_m}\cr
	&\ \ \ +(t-t_0)A(t_0) e^{-iE_m t}u_m(x)\int dx'\ u_m^*(x') V u_m(x') + (\mathrm{initial\ value})~.
\ealigned{}
\end{align}
We will choose an initial condition for $\psi_1$ to vanish at $t=t_0$. After choosing an appropriate initial condition, the solution of Eq.~\eqref{1} up to first order in $\epsilon$ is given by
\begin{align}
\baligned{}
\label{firstsol}
	\psi &= A(t_0)e^{-i E_m t}u_m(x) \cr
	&\ \ \ + \epsilon \left[ A(t_0)\sum_{n\neq m}u_n(x)\left(\int dx'u^*_n(x')Vu_m(x')\right) e^{-iE_n t}\frac{e^{i(E_n-E_m)t}}{E_n - E_m}\right.\cr
	&\ \ \ \ \ \ \ \left.-i(t-t_0)A(t_0) e^{-iE_m t}u_m(x)\int dx'\ u_m^*(x') V u_m(x') \right]~.
\ealigned{}
\end{align}
Now, we can impose the RG equation
\begin{align}\label{RGsch}
	\PDD{\psi}{t_0} = 0~,
\end{align}
to the expression~\eqref{firstsol} to obtain
\begin{align}
	\PDD{A}{t_0} + i \epsilon A(t_0)\int dx'\ u_m^*(x') V u_m(x') = 0~.
\end{align}
Solving this equation gives
\begin{equation}
\label{eq:A10}
	A(t_0) = A_0 e^{-i \epsilon \delta E_m t}~,\\
\end{equation}
with
\begin{equation}
\label{eq:A11}
	\delta E_m = \int dx'\ u_m^*(x') V u_m(x')~.
\end{equation}
Substituting Eq.~\eqref{eq:A10} and \eqref{eq:A11} into Eq.~\eqref{firstsol}, and setting the arbitrary reference time $t_0$ to $t$, we obtain the renormalized solution
\begin{align}
\baligned{}
	\psi &= A_0 e^{-i (E_m+\delta E_m) t}u_m(x) \cr
	&\ \ \ + \epsilon A_0\left[ \sum_{n\neq m}u_n(x)\left(\int dx'u^*_n(x')Vu_m(x')\right) e^{-iE_n t}\frac{e^{i(E_n-E_m)t}}{E_n - E_m}\right]~.
\ealigned{}
\end{align}
This reproduces well-known result for  the time independent perturbation theory of the Schr\"odinger equation. The essential point is that we have eliminated the secularly growing term in Eq.~\eqref{firstsol} by imposing RG equation \eqref{RGsch}. We apply the exactly same thing to the perturbative analysis of an axion cloud, but identification of the term to be renormalized is slightly different. The next example may clarify this point.

\subsection{Non-linear unstable oscillator}

Next, we will consider the following unstable oscillator
\begin{align}
	\label{unstableos}
	\left(\OOD{t} - 2 a \OD{t} + 1\right)x = - \epsilon x^2\ODD{x}{t}~,
\end{align}
with $0<a<1$ and $\epsilon \ll 1$. Note that this oscillator has a similar behavior to the superradiant instability of the axion cloud. The term $-2 a \ODD{x}{t}$ in the left hand side represents anti-dissipation, which  makes the oscillator unstable. On the other hand, the right hand side represents the dissipative effect, which corresponds to the radiative energy loss of the axion cloud.

We will solve Eq.~\eqref{unstableos} using perturbation theory. We expand the solution $x(t)$ as $x = x_0 + \epsilon x_1 + \cdots$. Substituting this expression to Eq.~\eqref{unstableos}, we get
\begin{align}\label{zerothosc}
	\left(\OOD{t} - 2 a \OD{t} + 1\right)x_0 &= 0~,
\end{align}
as the zeroth order equation. The solution to this equation is 
\begin{align}
\label{zerothordapp}
	x_0 = e^{a t}\left(C(t_0) e^{-i\omega t} + C(t_0)^* e^{i\omega t}\right)\,,
\end{align}
with $\omega = \sqrt{1 - a^2}$. We leave the integration constant $C(t_0)$ arbitrary in order to apply RG method. 

The first order equation is
\begin{align}
\baligned{}
	\left(\OOD{t} - 2 a \OD{t} + 1\right)x_1 &= - \left(C^3 (a - i \omega)e^{-3i \omega t} + |C|^2 C (3a - i \omega) e^{-i\omega t}\right.\cr
	&\ \ \ \ \ \ \left. + |C|^2 C^* (3a + i\omega)e^{i\omega t} + C^{*3} (a + i\omega)e^{3 i \omega t}\right)~,
\ealigned{}
\end{align}
which has the following solution:
\begin{align}
\label{firstordapp}
\baligned{}
	x_1 &= \frac{1}{2 i \omega}\left(\frac{C}{2}e^{2 a t}e^{-2 i \omega t} + \frac{3a - i \omega}{2a}C|C|^2e^{2at}+C^*|C|^2 \frac{3 a + i\omega}{2(a+i\omega)}e^{2 a t}2^{2i\omega t}\right.\cr
	&\ \ \ \left.+ C^{*3}\frac{a+ i \omega}{2(a + 2 i \omega)}e^{3 at}e^{4i\omega t}\right)e^{a t}e^{-i\omega t}+ \mathrm{c.c.} \cr
	& \ \ \ \ \ \ \  + C_1(t_0)e^{at}e^{- i \omega t}+ C_1^*(t_0)e^{at}e^{i \omega t}~.
\ealigned{}
\end{align}
The terms with $C_1(t_0)$ represent the freedom of adding homogeneous solutions to the particular perturbative solution. We observe that some terms behave as $\mathcal{O}(a^{-1})$ in the small $a$ limit, and thus breaks the validity of the perturbation. 

The term diverging in $a \to 0$ limit in Eq.~\eqref{firstordapp} becomes a secular term, which is proportional to $t-t_0$, after subtraction of an appropriate counterterm as
\begin{align}
	\frac{3a - i \omega}{4i\omega a}C|C|^2(e^{2at} - e^{2 a t_0})e^{a t}e^{-i\omega t} \to \frac{3a - i \omega}{2i\omega}C|C|^2 ( t- t_0)e^{a t}e^{-i\omega t}~,
\end{align}
which corresponds to choosing $C_1$ as 
\begin{align}
	C_1(t_0) = - \frac{3a - i \omega}{4 i \omega a}e^{2 a t_0}C|C|^2~.
\end{align} 
As in the case of the previous example, the problematic term can be renormalized.

Now, we apply the RG equation
\begin{align}
	\PDD{x}{t_0} = 0~,
\end{align}
to perturbative solution $x = x_0 + \epsilon x_1 + \dots$. After some calculations, we can derive an amplitude equation
\begin{align}
	\ODD{C}{t_0} =  \epsilon \frac{3a - i \omega}{2 i \omega}e^{2 a t_0}|C|^2 C~,
\end{align}
or
\begin{align}\label{ampeqosc}
	\ODD{|C|}{t_0} = - \frac{1}{2}\epsilon e^{2 a t_0}|C|^3~, 
	\qquad \ODD{\Theta}{t_0} = -\frac{3 a}{2 \omega}\epsilon e^{ 2 a t}|C|^2~,
\end{align}
if we write $C = |C|e^{i \Theta}$. Solutions to amplitude equations \eqref{ampeqosc} are
\begin{align}
	\label{ampapp}
	|C(t)|^2 &= \frac{C_0^2}{1 + \frac{\epsilon}{2a}C_0^2 (e^{2 a t}-1)}~,\\
	\label{phaseapp}
	\Theta(t) &=  -\frac{3 a^2}{2 \omega}\log\left(1 + \frac{\epsilon}{2 a}C_0^2(e^{2 a t}-1)\right)+\Theta_0~.
\end{align}
Here, $C_0,\Theta_0$ are the amplitude and the phase determined by the initial conditions of $x$. After substituting Eqs.~\eqref{ampapp} and \eqref{phaseapp} to Eqs.~\eqref{zerothordapp} and \eqref{firstordapp}, we obtain the renormalized first order solution to Eq.~\eqref{unstableos} as 
\begin{align}
\label{renomsolapp}
\baligned{}
	x &= 2 |C| e^{a t}\cos(\omega t - \Theta) \cr
	&\ \ \ + \frac{\epsilon}{2 (a^2 + 4 \omega^2)}|C|^3 e^{3 a t}\left(-  a \cos (3(\omega t - \Theta)) + 2 \omega \sin(3(\omega t - \Theta))\right)\cr
	&\ \ \ \ \ + \frac{\epsilon}{2w}|C|^3 e^{3 a t}\left(-2a\omega \cos(\omega t - \Theta) + (2 a^2+1)\sin(\omega t - \Theta)\right)~.
\ealigned{}
\end{align}

\begin{figure}[t]
	\centering
	\includegraphics[keepaspectratio,scale=0.7]{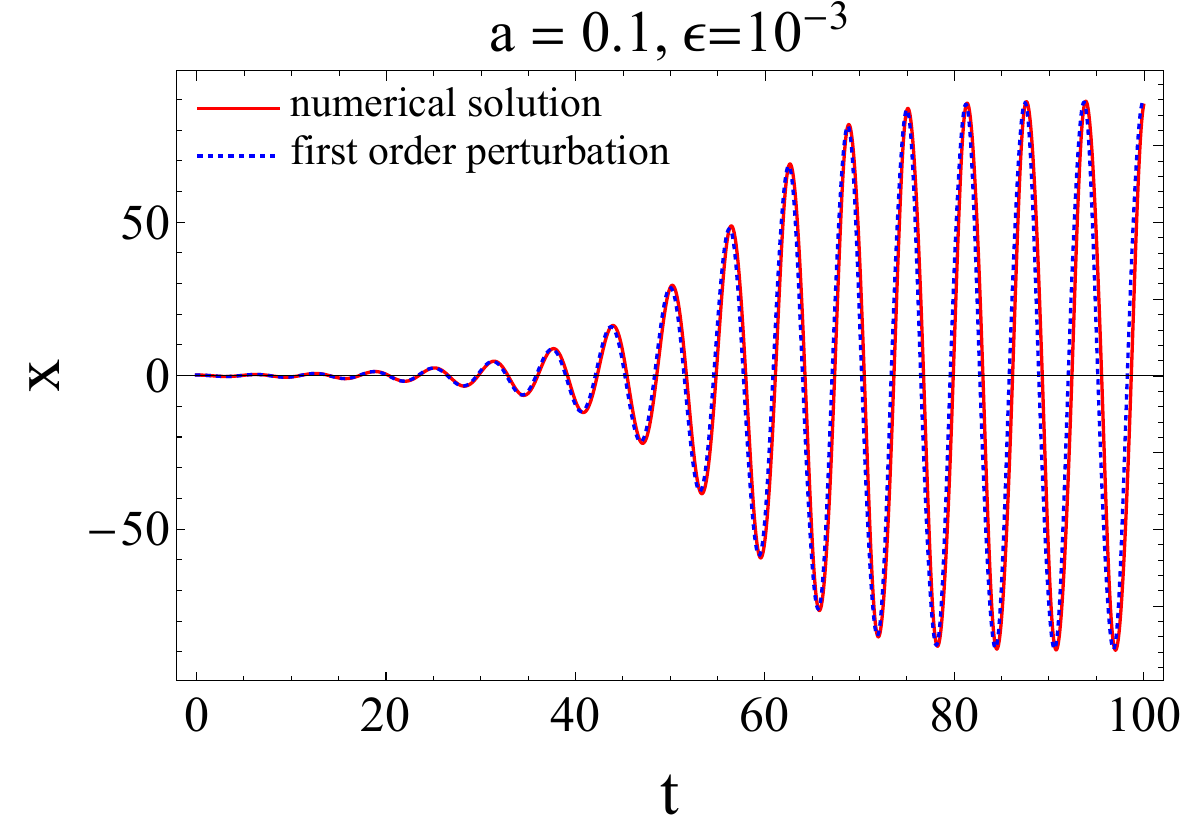}
	\caption{The red and blue dotted lines show numerical solution and renormalized solution of Eq.~\eqref{unstableos}, respectively. We took parameters as $a=0.1$ and $\epsilon =10^{-3}$.}
	\label{fig1app}
\end{figure}

To see how good our RG method are, we compared numerical solution of Eq.~\eqref{unstableos} and our renormalized solution Eq.~\eqref{renomsolapp}. Figure \ref{fig1app} shows the renormalized solution gives a good approximation. From this example, we believe that when the equation has a solution with a stable final state for which the perturbative correction remains to be small, we can obtain such a solution by the RG method, even though the zeroth order solution is unstable. 

Let us summarize the recipes of the RG method to solve a differential equation. First, obtain a naive perturbative solution, leaving integration constants of the zeroth order solution arbitrary. The problematic terms, which have a secular growth or are parametrically enhanced to violate the perturbative expansion, will be absorbed by an appropriate choice of the integration constants. Then, choose initial conditions for higher order solutions to correctly cancel the problematic terms in the perturbative solution at some chosen time $t_0$. Now, demand that the naive perturbative solution satisfies the RG equation. We can put $t_0 = t$ because solution no longer depends on $t_0$ due to the RG equation. After all, we obtain the solution without divergence.

\section{Scheme dependence of the amplitude equation}
\label{App:B}

In this appendix, we show that the ambiguity of the amplitude equation \eqref{eq:65} is related to the ambiguity in the definition of the cloud amplitude.

In general, amplitude equation takes following form: 
\begin{align}\label{appb:1}
	\ODD{A(\tau)}{\tau} = c_0 A(\tau) + c_1\lambda A(\tau)|A(\tau)|^2 + c_2 \lambda^2 A(\tau) |A(\tau)|^4 + \dots ~.
\end{align}
Here, constants $c_i\ (i=0,1,2,\dots)$ are determined from the renormalization. We can derive the amplitude equation for the different amplitude $\tilde{A}(\tau)$ as 
\begin{align}\label{appb:2}
	\ODD{\tilde{A}(\tau)}{\tau} = \tilde{c}_0 \tilde{A}(\tau) + \tilde{c}_1\lambda \tilde{A}(\tau)|\tilde{A}(\tau)|^2 + \tilde{c}_2 \lambda^2 \tilde{A}(\tau)|\tilde{A}(\tau)|^4 + \dots~.
\end{align}
Now, assume that these two amplitudes are related by the following equation:
\begin{align}\label{appb:3}
	\tilde{A}(\tau) = A(\tau) + a_1 \lambda A(\tau)|A(\tau)|^2 + a_2 \lambda^2 A(\tau)|A(\tau)|^4 + \dots~,
\end{align}
Substituting Eq.~\eqref{appb:3} into Eq.~\eqref{appb:2} and using Eq.~\eqref{appb:1}, we obtain
\begin{align}\label{appb:4}
	\tilde{c}_0 &= c_0~,\\
	\label{appb:5}
	\tilde{c}_1 &= c_1 + a_1 (c_0 + c_0^*)~,\\
	\label{appb:6}
	\tilde{c}_2 &= c_2 + (2 a_1^2 - |a_1|^2 + 2 a_2)(c_0 + c_0^*) + a_1 c_1^* - c_1 a_1^*~.
\end{align}
This shows that the ambiguity in the choice of the homogeneous solution in RG method, which is shift in $c_i$, is related to the definition of the amplitude. This result is general for any flow equation, such as rernomalization group flow in the QFT. In this case, amplitude corresponds to the coupling constant.

Now we concentrate on the axion cloud case. In this case, we know that $c_0 = \tilde{c_0} = \omega_{0,I}$ (see Eq.~\eqref{eq:65}). Therefore, as long as relation between two amplitudes \eqref{appb:3} does not contain $\mathcal{O}(\omega_{0,I}^{-n}),~ n=1,2,\dots$ quantities, difference in the $c_1$ and $\tilde{c}_1$ is suppressed by $\omega_{0,I}$. We can then identify $a_1 = \delta C^{(1)}$. This shows the ambiguity in the counterterm is the ambiguity in the definition of the amplitude.

Now we look at the second order equation. Eq.~\eqref{appb:6} shows that difference in real part of $c_2$ is always suppressed by $\omega_{0,I}$. Therefore, adjusting the amplitude by first order in $\lambda$ won't eliminate the divergence in the second order solution.

\section{Derivation of the dissipative effect $C^{(2)\diss}_{l_0m_0\omega_0}$}
\label{sec:AppC}

In this appendix, we derive Eq.~\eqref{eq:C2diss}. We start with the evolution equation of the particle number and convert this to that of the amplitude, then identify the coefficient which correspond to the $C^{(2)}$ in the RG equation.

The number density current $J_\mu$ for the mode $\Psi = \mathcal{A} e^{-i(\omega t - m \varphi)}R_{lm\omega} S_{lm\omega}$ is given by
\begin{align}
    J_{\mu} = -i \left(\Psi^* \partial_\mu \Psi - \Psi \partial_\mu \Psi^*\right)
\end{align}
and obeys conservation equation
\begin{align}
    \nabla_\mu J^\mu = 0~.
\end{align}
Following the derivation of Eq.~\eqref{energyintegral}, we obtain
	\begin{gather}\label{eq:C3}
		\ODD{N}{t} = - F~,\\
		N = \int_{\Sigma} dr d\theta  d\phi \sqrt{- g}J^{t}~,\\
		\label{eq:C5}
		F = - \int_{\partial\Sigma} d\theta \int d\phi \sqrt{- g}J^{r}~,\\
		J^r = - i \frac{\Delta}{\rho^2}\left(R_{lm\omega}^*\partial_r R_{lm\omega} - R_{lm\omega} \partial_r R_{lm\omega}^*\right) |S_{lm\omega}|^2 |\mathcal{A}|^2~.
	\end{gather}
Here, $\Sigma$ is constant $t$ surface and $\partial \Sigma$ is the boundary of $\Sigma$, which is $r = r_+$ and $r = \infty$. Therefore, $N$ is the particle number on the constant $t$ surface and the $F$ is the flux through the boundary. 

When we take superradiant mode $(l_0,m_0,\omega_0)$ for $\Psi$ and neglecting the self-interaction, then the the right hand side of Eq.~\eqref{eq:C3} is given by
\begin{align}
    F = 2\pi i \left(\int d\theta |S_{l_0m_0\omega_0}|^2 \right) \Delta \left(R_{l_0m_0\omega_0}^*\partial_r R_{l_0m_0\omega_0} - R_{l_0m_0\omega_0} \partial_r R_{l_0m_0\omega_0}^*\right)|_{r = r_+} |\mathcal{A}|^2~.
\end{align}
There is no contribution from the infinity due to the decaying boundary condition. Using the boundary condition \eqref{inhorizon}, we obtain
\begin{align}
    \Delta \left(R_{l_0m_0\omega_0}^*\partial_r R_{l_0m_0\omega_0} - R_{l_0m_0\omega_0} \partial_r R_{l_0m_0\omega_0}^*\right)|_{r = r_+} = 2 i (\omega - m \Omega_H) (r_+^2 + a^2)~.
\end{align}
Thus, $F$ is given by
\begin{align}\label{eq:C9}
    F = 4 \pi (\omega - m \Omega_H) (r_+^2 + a^2) \left(\int d\theta |S_{lm\omega}|^2 \right) |\mathcal{A}|^2~.
\end{align}
On the other hand, left hand side of Eq.~\eqref{eq:C3} is given by
\begin{align}\label{eq:C10}
    \ODD{N}{t} = 2 \omega_I N~,
\end{align}
because the frequency $\omega$ is complex. Substituting Eqs.~\eqref{eq:C9} and \eqref{eq:C10} into Eq.~\eqref{eq:C3}, we obtain the relation between the particle number and the amplitude as
\begin{align}\label{eq:C11}
		 N = - 2 \pi \omega_I^{-1} (\omega - m \Omega_H) (r_+^2 + a^2) \left(\int d\theta |S_{lm\omega}|^2 \right) |\mathcal{A}|^2~.
\end{align}

Now we include the effect of the self-interaction by evaluating the flux \eqref{eq:C5} of the first order solution \eqref{renomfirstorder}. The first order solution has two contributions to the flux. One is dissipation into infinity through $(l,3m_0,3\omega_0)$ modes and other is accumulation via superradiant scattering by the rotating black hole. Because superradiant scattering is suppressed by tunneling through angular momentum barrier, we only consider the disipation into infinity. The dissipative part in the first order solution is given by
\begin{align}
\baligned{}
	\phi_{(1)}(x) \supset&\, e^{-3i(\omega_0 t - m_0 \varphi)}\sum_{l} S_{l3m_03\omega_0}(\theta)
	\int dr'd\cos\theta'\ (r'^2 + a^2 \cos^2\theta')
	\cr & \qquad \times S_{l3m_03\omega_0}(\theta')G^{3\omega_0}_{l3m_0}(r,r')S_{l_0m_0\omega_0}(\theta')^3 R_{l_0m_0\omega_0}(r')^3~.
\ealigned{}
\end{align}
Using the boundary condition Eq.~\eqref{outinf}, behavior around infinity is
\begin{align}
\baligned{}
	\phi_{(1)} \to &\, A^3 e^{-3i(\omega_0 t - m_0 \varphi)}\,\sum_{l} \frac{1}{W_{l3m_0}(3\omega_0)} S_{l3m_03\omega_0}(\theta) \frac{e^{+i\sqrt{9\omega_0^2-\mu^2}r_*}}{r} \cr
	& \times \int dr'd\cos\theta'\ (r'^2 + a^2 \cos^2\theta')S_{l3m_03\omega_0}(\theta')R^{r_+}_{l3m_03\omega_0}(r')S_{l_0m_0\omega_0}(\theta')^3 R_{l_0m_0\omega_0}(r')^3
\ealigned{}\cr
  &  \equiv  A^3 e^{-3i(\omega_0 t - m_0 \varphi)}\sum_{l} j_{l3m_03\omega_0} S_{l3m_03\omega_0}(\theta) \frac{e^{+i\sqrt{9\omega_0^2-\mu^2}r_*}}{r}~.
\end{align}
Substitute this into Eq.~\eqref{eq:C5}, we obtain the flux of the first order solution $F_{(1)}$ as 
\begin{align}\label{eq:C14}
    F_{(1)} = 2\sqrt{9\omega_0^2 - \mu^2} |j_{l3m_03\omega_0}|^2 |A|^6~.
\end{align}

From Eqs. \eqref{eq:C3}, \eqref{eq:C10} and \eqref{eq:C14}, time development of the particle number is given by
\begin{align}
    \ODD{N}{t} = 2 \omega_{0,I} N - 2\sqrt{9\omega_0^2 - \mu^2} |j_{l3m_03\omega_0}|^2 |A|^6~.
\end{align}
Using Eq.~\eqref{eq:C11}, this is converted to the time evolution of the amplitude:
\begin{align}\label{eq:16}
    \frac{1}{\omega_{0,I}}\ODD{|A|}{t} =  |A| + \frac{\sqrt{9\omega_0^2 - \mu^2}\sum_{l}|j_{l3m_03\omega_0}|^2}{2 \pi (\omega_0 - m_0 \Omega_H) (r_+^2 + a^2) \left(\int d\theta |S_{l_0m_0\omega_0}|^2 \right)}|A|^5~.
\end{align}
The coefficient of the $|A|^5$ term represent the dissipative effect due to the self-interaction on the evolution of the amplitude. Therefore, we define
\begin{align}
    C^{(2)\diss} = \frac{\sqrt{9\omega_0^2 - \mu^2}\sum_{l}|j_{l3m_03\omega_0}|^2}{2 \pi (\omega_0 - m_0 \Omega_H) (r_+^2 + a^2) \left(\int d\theta |S_{l_0m_0\omega_0}|^2 \right)}~.
\end{align}
$C^{(2)\diss}$ is the leading dissipative part of the $\Re{C^{(2)}}$, because any excited modes other then $(l,3m_0,3\omega_0)$ modes dissipate energy (see discussion below Eq.~\eqref{defC2}).
\bibliographystyle{ptephy}
\bibliography{axionref}
%
\end{document}